\documentclass[twocolumn,trackchanges]{aastex7}
\usepackage{subcaption} 
\usepackage{tabularx}
\usepackage{amsmath}
\usepackage{xfrac}

\newcommand{\chandra}{{\it Chandra}}
\newcommand{\planck}{{\it Planck}}
\begin{document}

\title{Clusters Hiding Under Millimeter Sources (CHUMS) I: Extreme CHUMS }

\author[orcid=0009-0006-8186-7440]{Harshda Saxena}
\affiliation{California Institute of Technology, 1200 East California Boulevard, Pasadena, California, USA}
\email[show]{hsaxena@caltech.edu}  

\author[orcid=0000-0002-8031-1217]{Adam B. Mantz}
\affiliation{Kavli Institute for Particle Astrophysics and Cosmology, Stanford University, 452 Lomita Mall, Stanford, CA 94305, USA}
\email{amantz@stanford.edu}

\author[orcid=0000-0002-8213-3784]{Jack Sayers}
\affiliation{California Institute of Technology, 1200 East California Boulevard, Pasadena, California, USA}
\email{jsayers@astro.caltech.edu}

\author[orcid=0009-0008-0072-120X]{Denise G. Yudovich}
\affiliation{Department of Astronomy, University of Florida, 211 Bryant Space Science Center, P.O. Box 112055, Gainesville, FL 32611, USA}
\affiliation{Department of Astrophysical \& Planetary Sciences, University of Colorado, 2000 Colorado Ave, Boulder, CO 80309, USA}
\email{deniseyudo@gmail.com}

\author[orcid=0000-0003-0667-5941]{Steve W. Allen}
\affiliation{Kavli Institute for Particle Astrophysics and Cosmology, Stanford University, 452 Lomita Mall, Stanford, CA 94305, USA}
\affiliation{Department of Physics, Stanford University, 382 Via Pueblo Mall, Stanford, CA 94305, USA}
\affiliation{SLAC National Accelerator Laboratory, 2575 Sand Hill Road, Menlo Park, CA  94025, USA}
\email{swa@stanford.edu}

\begin{abstract}
Galaxy cluster abundance provides a powerful probe of the $\Lambda$CDM model and enables precise constraints on cosmological parameters. Millimeter-wavelength surveys detect clusters through the Sunyaev–Zeldovich (SZ) effect, and are particularly effective at high redshifts. However, the SZ signal can be significantly contaminated by emission from Active Galactic Nuclei (AGN), particularly AGN within the Central Galaxies (CGs). This contamination reduces the SZ signal strength at the frequencies most accessible from the ground, which reduces detection significances or converts cluster detections to non-detections, thereby diminishing survey completeness and introducing biases in cosmological analyses. In this work, we analyze three clusters that host bright AGN in their CGs using 30 and 90~GHz observations from the Combined Array for Research in Millimeter-wave Astronomy (CARMA). In each case the AGN emission overwhelms the cluster SZ signal, resulting in non-detections in the Atacama Cosmology Telescope (ACT) survey. We present signal to noise ratio (SNR) estimates for the clusters after subtracting the AGN signal from 90~GHz ACT maps using the CARMA measurements, demonstrating high SNR cluster detections once this contaminating emission is removed. Using cluster pressure profiles derived from \chandra\ X-ray data, we subtract the expected SZ signal from the 150~GHz ACT maps to estimate the flux density of the AGN in that band. Leveraging the time-asynchronous CARMA observations, we also assess temporal variability in the AGN emission, and find low fractional variability for our sample. Finally, we discuss the importance of modeling and mitigating AGN contamination in SZ cluster surveys. 

\end{abstract}



\section{Introduction} \label{sec:intro}
As the most massive gravitationally bound structures in the Universe, galaxy clusters trace the late stages of structure formation, representing the evolution of the highest peaks in the primordial matter density field. Their abundance is particularly useful for constraining both the mean matter density ($\Omega_m$) and the amplitude of matter fluctuations ($\sigma_8$; \citealt{Allen2011}). Studies using cluster catalogs selected from X-ray, optical, and millimeter-wavelength surveys have yielded precise constraints on these parameters \citep{Mantz1407.4516, Planck1502.01597, Lesci_2022,Bocquet2024,Ghirardini_2024,DESY3_2025_Halo_Abundance} but have been systematically limited by uncertainties related to halo mass calibration and selection biases \citep{von_der_Linden_2014, Zubeldia_2019, Willis_Xrayselbias, Lesci_2023, Optical_selbias}.
\\\\
Wide-area surveys detect clusters based on observables,  such as the total integrated Sunyaev-Zel’dovich (SZ) effect signal \citep{Bleem1409.0850, Planck1502.01598, Hilton2009.11043}, X-ray emission \citep{Ebeling_1998,Ebeling0009101, Bohringer_2004, Bulbul2402.08452}, and optical richness \citep{Rykoff1303.3562, Rykoff1601.00621, DES2002.11124}. The survey selection function, representing the probability of detecting a cluster as a function of its intrinsic properties \citep{Melin_selection, Allen2011, Mantz1901.10522}, is a critical ingredient that must be accurately quantified for precise cosmological parameter estimation. Mm-wavelength surveys selected using the SZ effect are inefficient at low masses, but provide an effective route to finding massive clusters out to $z \gtrsim 2$, due to the SZ effect surface brightness being independent of redshift \citep{Carlstrom_2002, Bleem_2024, ACT_DR6_clusters}. Contamination of the SZ signal by radio-bright galaxies within clusters is a well-established systematic in SZ cluster detection \citep{Gralla_2014, Gupta2017}. However, its overall impact remains poorly quantified, despite the existence of several striking individual cases of contamination \citep{Coble_2007,Bonamente_2012,Sayers1209.5129, Hogan15b}.
\\\\
Active galactic nuclei (AGN) hosted by the cluster in their central galaxies (CGs) play a critical role in the thermodynamic and dynamical evolution of galaxy clusters \citep{Fabian_2012, McNamara_2012, Werner_2018}. Mechanical feedback from the central AGN is thought to prevent runaway cooling \citep{Fabian_2012}, provide non-thermal pressure through turbulent motions \citep{Zhuravleva1410.6485, Zhuravleva1707.02304, Irina_Mach, Hitomi_perseus,Vazza,Eckert_2019,Heinrich_2024, Dupourqu_2024}, and quench star formation in the central galaxies \citep{Joop_AGN, AGN_quench_SF, SF_met_tracing}. Observational evidence for this mechanical feedback is seen through the presence of X-ray cavities in the intracluster medium, coincident with radio emission \citep{Fabian_2000,Fabian_2003,McNamara_2000,Forman_2005,Forman_2007,Fabian_2012,Hlavacek_Larrondo_2012, Pandge_2021}. The SZ effect leads to a negative signal in the millimeter bands, whereas the AGN emission is positive. The typical angular resolution ($\gtrsim 1\arcmin{}$) for SZ surveys makes it difficult to spatially separate the SZ and AGN signals. Thus, the associated synchrotron emission from radio-bright AGNs can significantly degrade SZ-based cluster detection by partially or fully masking the SZ decrement. This may lead to reduced detection significance or complete non-detections, introducing incompleteness that directly biases measurements of the amplitude and growth of cosmic structure inferred from SZ surveys. \cite{Gupta2017} find that, depending on the redshift evolution of the radio galaxy luminosity function, 1.8--5.6\% of clusters are lost from the sample in an SPT like survey. Further, \cite{Mo_AGN_mass_clusters} show that that SZ-based cluster masses with mm-bright AGNs are expected to be biased low by at least 20\% for $1-7\%$ of clusters, and approximately half of those are biased low by 50\% or more. Although high angular resolution observations are essential for identifying and characterizing this contamination, most existing studies are limited to radio frequencies $\leq 1.4$ GHz, leaving the emission from CGs in the relevant $\sim$90–300 GHz millimeter regime poorly constrained \citep{Hardcastle0804.3369, Hogan15b}. The greater spectral coverage of Planck partially mitigated this issue in its SZ cluster survey, however its limited spatial resolution reduced its discovery potential at the highest redshifts. Upcoming and active ground-based SZ surveys with sufficient sensitivity to detect higher redshift clusters, but with course angular resolution and limited observing bands, must account for this contamination.
\\\\
Here we analyze three galaxy clusters identified through a visual inspection of 90\,GHz Combined Array for Research in Millimeter-wave Astronomy (CARMA) observations of galaxy clusters, selecting systems that host extremely millimeter-bright central AGN. All three of these clusters are missing from the ACT cluster catalog, but two are detected in \planck\ due to its greater spectral coverage. The single non-detection by \planck\ appears to be due to emission from a strong IRAS source nearby which is confused with the cluster and AGN at Planck's spatial resolution. 

The first cluster, MACSJ0242.5–2132 (hereafter MACS0242), is a cool-core, dynamically relaxed system at a redshift of 0.314 \citep{Allen_2004, Ebeling_2010}. The CG in MACS0242 is a luminous radio AGN, expected to be a low-peaking Gigahertz-Peaked Spectrum (GPS) source with a spectral turnover below 1 GHz \citep{Condon_1978, Hogan15b}. Very Long Baseline Array (VLBA) observations reveal a diffuse, “wispy” morphology on parsec scales, consistent with the source being in a late stage of nuclear activity \citep{Hogan15b}. 

The second system, RX J0439.0+0520 (hereafter RXJ0439), is a massive, relaxed cluster at a redshift of 0.208 that hosts a bright, spatially extended radio source \citep{Coble_2007,Allen_2004, Allen_2007}. Its spectral energy distribution (SED) is well described by a strong GPS component accompanied by a steep-spectrum power law at lower frequencies, indicative of recent powerful activity in the cluster core \citep{Hogan15b}. Below the self-absorption turnover of the core component, the spectrum is dominated by an underlying steep power-law emission, which may point to the presence of an amorphous halo of confused radio emission \citep{Hogan15b}. 

The final system we consider is the optically poor cluster RXJ1651.1+0459 (hereafter RXJ1651) hosting the AGN referred to as Hercules A at a redshift of 0.155 \citep{Hart_RXJ1651}, which is an exceptionally luminous radio source. Hercules A exhibits highly complex radio morphology that does not conform cleanly to the standard Fanaroff–Riley (FR) I or II classifications. Its radio lobes possess sharp outer edges rather than fading gradually into the intracluster medium \citep{Timmerman_2023}. The eastern lobe is dominated by a bright jet that progressively disperses into the lobe, while the western lobe is characterized by a striking sequence of three rings, each examined in detail by \cite{Timmerman_2023}. Owing to this morphological complexity, modeling Hercules A presents substantial challenges; consequently, we present only limited results for this system.
\\\\
This paper is organized as follows. In Section \ref{data}, we describe the CARMA data reduction and analysis procedures and provide details of the Multi Matched Filtering (MMF3) pipeline used to quantify cluster detection probabilities. In Section \ref{result}, we present our measurements of the AGN flux densities at millimeter wavelengths and present their Spectral Energy Distribution (SED) in this regime, examine temporal variability in the radio emission, and describe our measurements for Hercules A. Finally, in Section \ref{conc} we discuss the importance of modeling and mitigating AGN contamination in SZ-selected cluster surveys to reduce systematic biases in cluster detection and mass estimation.

\section{Data collection and analysis} \label{data}
\subsection{CARMA data}
CARMA was a heterogeneous interferometric array comprised of six 10.4 m, nine 6.1 m, and eight 3.5 m antennas. In this work, we use only data acquired by the 8-element subarray of 3.5\,m CARMA antennas, processed by the 8\,GHz CARMA wideband correlator. Observations were made in the SH and SL configurations, which both feature 6 antennas with compact spacings and 2 outlying antennas to provide longer baselines. The observation dates, on-source integration times, observing frequencies, and flux calibrators for each cluster are summarized in Table \ref{tab:data} in Appendix \ref{targets_list}.
\\\\
Data reduction was carried out using {\sc miriad} \citep{Miriad}, following the procedure described by \cite{Muchovej_2007}. The data were flagged for poor weather, antenna shadowing, elevated system temperatures, and other technical issues. Bandpass and complex gain calibration were performed using observations of bright quasars interleaved with the target observations. Each observation also included measurements of a bright, unresolved planet—Mars, Uranus, or Neptune—which were used to establish the absolute flux scale, tied to {\it Planck} calibration \citep{Planck_calib_planet, Planck_planet, Maris_2021}. For each cluster and observing frequency, the RMS noise level, synthesized beam full width at half maximum (major and minor axes), beam position angle, and the uv coverage are listed in Table \ref{tab:data_radiorms_beam}. AGN positions and flux densities were fitted directly to the calibrated visibilities using {\sc difmap} \citep{difmap}, with $uv$ radii as given in the table, where the cluster SZ signal is comparatively negligible.

\begin{table*}[ht]
    \centering
    \caption{RMS noise level and synthesized FWHM beam widths for each cluster at each frequency}
    \label{tab:data_radiorms_beam}
    \begin{tabular}{|c|c|c|c|c|c|c|}
        \hline
        \textbf{Cluster Field} & Freq. (GHz) & $b_{\mathrm{min}}$ (arcsec) & $b_{\mathrm{maj}}$ (arcsec) & P.A (degrees) & RMS noise (mJy/beam) & UV coverage ($k\lambda$)\\
        \hline
        MACS0242 & 31.4 & 63.71 & 118.2  & -9.73 & 1.79 & 1.2 - 1.6\footnote{In this particular observation, both of the outrigger antennas were flagged, requiring us to use relatively small $uv$ radii than for the other measurements. Using the \chandra\ derived pressure profile discussed in Section \ref{result}, we estimate the contamination from the cluster's SZ signal at these baselines to be less than 0.46 mJy, negligible compared to the AGN emission. } \\ 
        \hline
        MACS0242 & 94.3 & 8.76 & 17.35 & -1.41 & 0.45 & 2.0 - 26.0 \\ 
        \hline
        RXJ0439 & 35.9 & 14.56 & 20.12 & 43.10 & 0.32 & 3.8 - 12.7 \\ 
        \hline
        RXJ0439 & 94.3 & 7.92 & 9.07 & 73.74 & 1.73 & 2.0 - 19.0 \\ 
        \hline
        RXJ1651 & 31.4 & 26.16 & 38.47 & 39.79 & 0.22 & 0.4 - 7.0 \\ 
        \hline
        RXJ1651 & 92.0 & 9.09 & 13.47 & 48.91 & 0.45 & 1.0 - 19.0 \\ 
        \hline
    \end{tabular}
\end{table*}

\subsection{MMF3 pipeline for SNR computation}
We independently reimplemented the MMF3 algorithm used by \cite{PSZ2}, with verification of our implementation described in \cite{Saxena_2025}. In brief, the total signal in the \planck\ maps at a given position in the sky is modeled as the combination of the SZ signal from a cluster parameterized by the generalized Navarro-Frenk-White (GNFW) profile \citep{Nagai_2007}, along with astrophysical and instrumental noise. A Multi-Matched Filter (MMF) is applied, which uses both spatial and frequency weighting to optimally extract the cluster signal from the maps. The noise matrix is estimated directly from these local maps, and when combined with the signal obtained above allows us to measure the cluster SNR. Our code was modified to accommodate ACT data in place of {\it Planck} observations. We use the ACT DR6 coadded night maps at 98 and 150 GHz, which have a pixel resolution of 0.5\arcmin{} \citep{ACT_DR6}. To estimate the expected SNR of each cluster, we inject mock clusters with cluster profiles inferred from a \chandra\ X-ray analysis (see Section \ref{sec:90results}) into random positions within the ACT survey. We avoid injecting these mock clusters within 1 Mpc of any cluster detection listed in the ACT DR6 cluster catalog \citep{ACT_DR6_clusters}. We optionally include the AGN as part of this injection, to estimate the cluster SNR with and without AGN contamination. These cluster and AGN profiles are convolved with the ACT beam presented in \cite{Choi_2020}, and the MMF3 algorithm is run on the resulting maps to generate a distribution of expected SNRs.

\section{Results} \label{result}
\subsection{Contamination of the SZ effect at 90 GHz}
\label{sec:90results}
\begin{figure*}
    \centering
    \includegraphics[width=\linewidth]{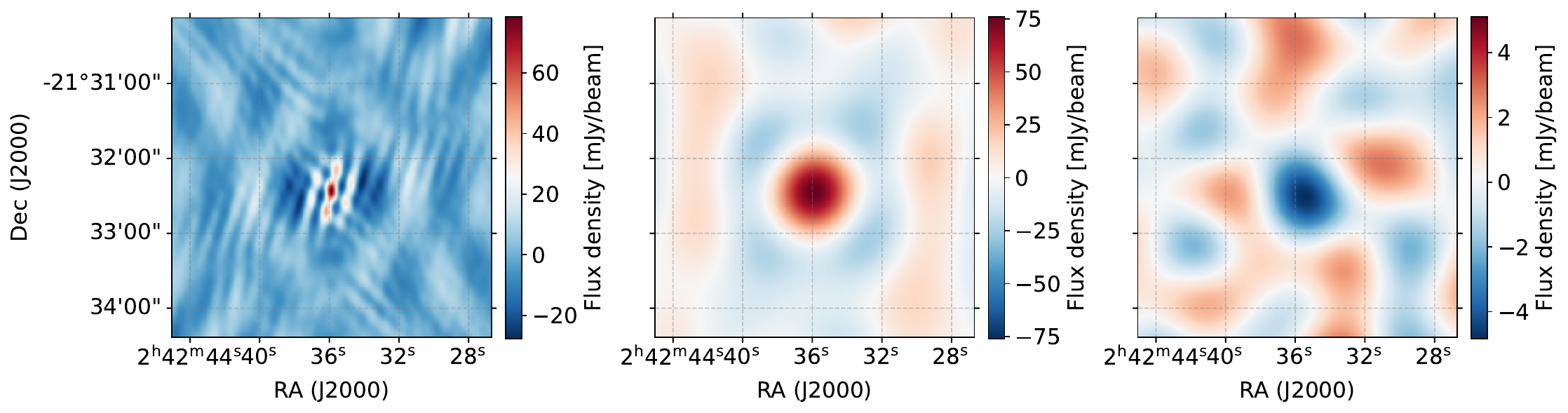}
    \caption{Left: CARMA data for long-baselines ($\geq 3k \lambda$) for MACS0242, dominated by AGN emission. Middle : CARMA short-baseline ($\leq 3 k\lambda$) data for MACS0242. Right: Residual CARMA short-baseline map, after subtracting the AGN emission and revealing the SZ decrement from the cluster.}
    \label{fig:macs_basesub}
\end{figure*}

\begin{figure*}
    \centering
    \includegraphics[width=\linewidth]{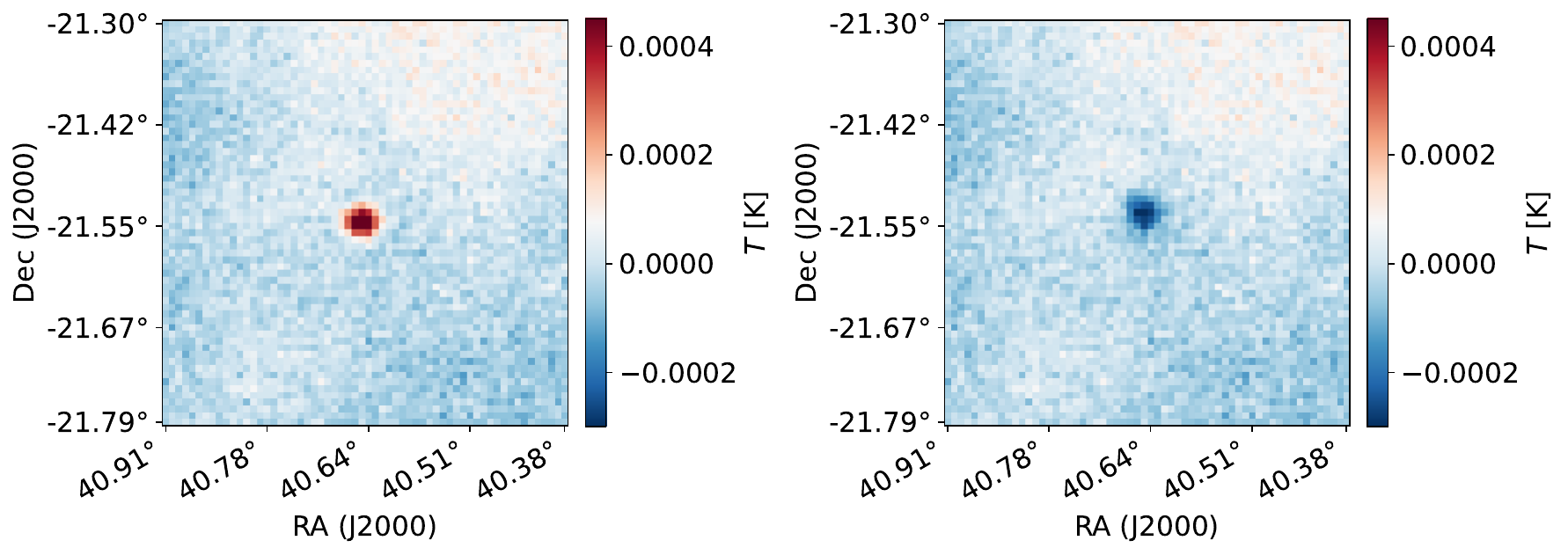}
    \caption{Left: ACT $98$ GHz map at the position of MACS0242. Right: ACT map after subtraction of the AGN emission, revealing the SZ decrement from the cluster.}
    \label{fig:macs_agnsub}
\end{figure*}

\begin{figure*}
        \centering
        \begin{subfigure}[t]{0.45\linewidth}
            \centering
            \includegraphics[width=\linewidth]{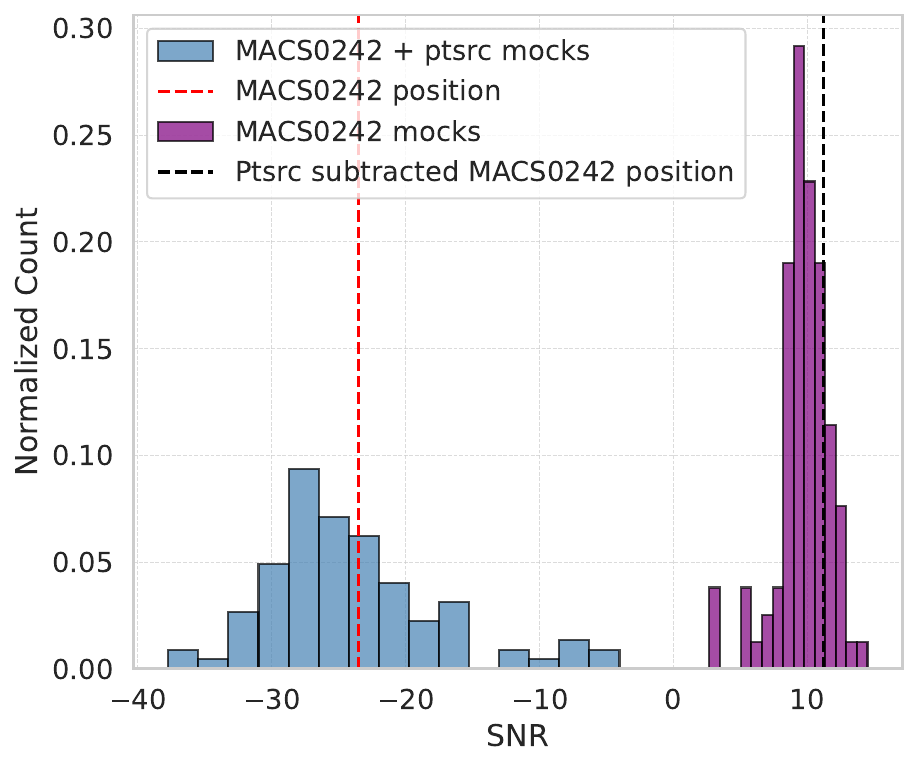}
            \caption{MACS0242}
            \label{fig:MACS0242_tot}
            \end{subfigure}
        \hfill
        \begin{subfigure}[t]{0.45\linewidth}
            \centering
            \includegraphics[width=\linewidth] {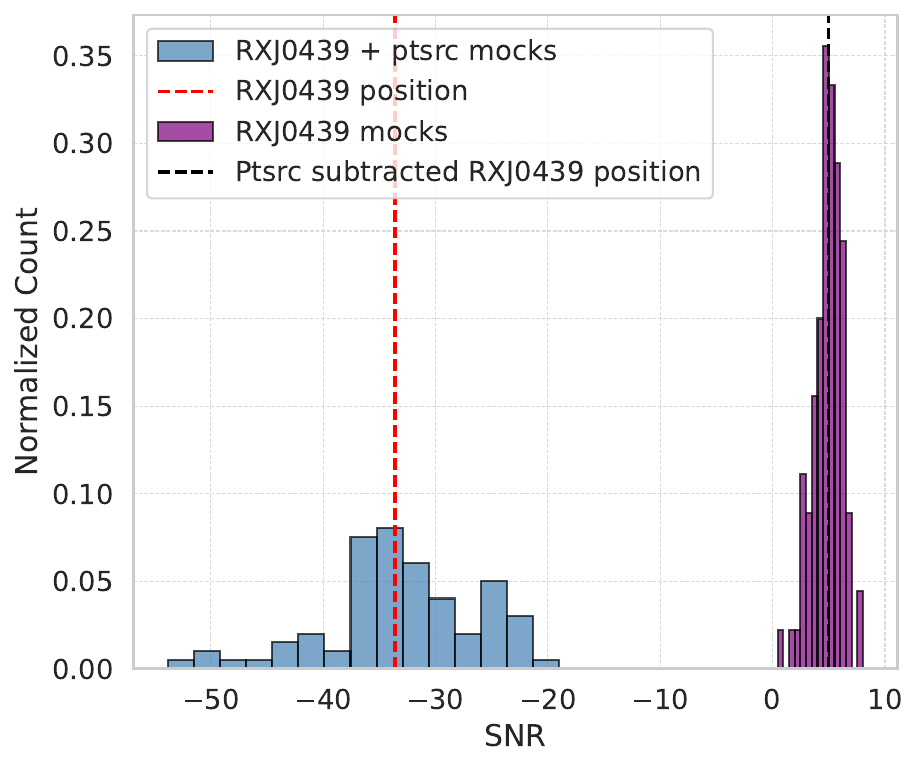}
            \caption{RXJ0439}
            \label{fig:RXJ0439_tot}
            \end{subfigure}
        \caption{
    Histograms in blue show the SNR distribution from mock observations of the cluster SZ signal based on the X-ray-derived pressure profile plus the bright central AGN, with the actual SNR from the location of the cluster shown in the red line. Histograms in purple show the SNR distribution from mock observations that include only the SZ signal (i.e., no central AGN), with the actual SNR from the location of the cluster following subtraction of the AGN emission shown in the black line.}
    \label{fig:combined_SNR}
\end{figure*}

Applying the CARMA analysis pipeline described in Section \ref{data} to MACS0242 and RXJ0439, we measure flux densities at the CARMA observing frequencies of $\sim30$ and 90~GHz, as summarized in Table \ref{tab:data_radioflux}. The CARMA data for long baselines, short baselines and short baselines after AGN subtraction are shown in Fig \ref{fig:macs_basesub}. These measurements are extrapolated to 98 GHz (corresponding to the central frequency of the lower ACT band) using a power law fit to the CARMA measurements. Based on the ACT beam profiles of \cite{Choi_2020}, the extrapolated flux densities are converted to brightness temperature units appropriate for ACT. To estimate the expected SZ signal in the ACT data, we utilize a fit to the cluster pressure profile measured from Chandra X-ray data by \cite{Mantz_2016}. In short, we fit for the normalization and scale radius of a GNFW profile, while fixing the inner and intermediate slopes to the values appropriate for cool-core clusters \citep{Mantz_2016} and the outer slope to the simulation-based value reported by \cite{Arnaud_2010}.
\\\\
To estimate the expected SNR in the ACT map based on the X-ray-derived cluster model and the measured AGN flux density, we draw 100 random noise cutouts from the unmasked regions described in Section \ref{data}. We create a mock observation from each cutout by injecting a cluster signal based on the GNFW model, along with a point-source model placed at the center of the cluster with brightness temperature as extrapolated from the CARMA measurements. The MMF3 pipeline is then applied to each realization, with the filter scale fixed to the expected angular extent of the cluster in order to maximize sensitivity to the intrinsic SZ signal. The resulting SNR distributions from these 100 mocks are shown in Fig. \ref{fig:combined_SNR}. For comparison, we also report the SNR measured at the true cluster position in the ACT maps, which is consistent with the distribution obtained from the mocks. This indicates that our AGN brightness measurement from CARMA are consistent with the X-ray-derived model of the SZ signal and the true signal observed by ACT. For both clusters, we find a large negative SNR, indicating a non-detection. 
\\\\
To estimate the intrinsic cluster signal, we create a new set of mock observations that contain only the GNFW model of the SZ signal -- i.e., no AGN emission -- and repeat the MMF3 analysis while varying the filter scale. In addition, we subtract the AGN emission from the observed ACT maps based on the measured positions and 98~GHz brightness values obtained from the CARMA data, and recompute the MMF3 SNR, also varying the filter scale. The resulting SNR values from the observed data are again consistent with those derived from the AGN-free mocks, as shown in Fig. \ref{fig:combined_SNR}. This validates our AGN subtraction procedure and yields high-significance SZ detections for both MACS0242 and RXJ0439. The ACT maps before and after AGN subtraction for MACS0242 are shown in Fig. \ref{fig:macs_agnsub}. 

\begin{table*}[ht]
    \centering
    \caption{Extracted flux densities for central AGN from CARMA, along with power-law extrapolations to higher frequencies}
    \label{tab:data_radioflux}
    \begin{tabular}{|c|c|c|c|}
        \hline
        \textbf{Cluster} & Flux density (mJy) & Frequency (GHz) & Source\\
        \hline
        MACS0242 & $176.6 \pm 1.9$ & 31.4 & CARMA \\
        \hline 
        MACS0242 & $78.1 \pm 2.6$ & 94.3 & CARMA \\
        \hline
        MACS0242 & $75.9 \pm 2.6$ & 98 & Extrapolation\\
        \hline
        MACS0242 & $55.3 \pm 2.6$ & 150 & Extrapolation \\
        \hline
        MACS0242 & $55.8 \pm 6.8$ & 150 & ACT map \\
        \hline
        RXJ0439 & $204.2 \pm 0.4$ & 35.9 & CARMA \\
        \hline 
        RXJ0439 & $96.2 \pm 0.9$ & 94.3 & CARMA \\
        \hline
        RXJ0439 & $93.4 \pm 1.0 $ & 98 & Extrapolation \\ 
        \hline
        RXJ0439 & $67.0 \pm 1.0 $ & 150 & Extrapolation \\ 
        \hline
        RXJ0439 & $64.0 \pm 7.6$ & 150 & ACT map \\ 
        \hline
    \end{tabular}
\end{table*}

\subsection{Contamination of the SZ effect at 150 GHz}\label{150_overest}
Based on the same power law fit to the CARMA flux density measurements at 30 and 90~GHz, we also estimate the AGN emission in the higher frequency ACT band centered at 150 GHz. Owing to the typical exponential steepening of the synchrotron spectrum from electron aging, this extrapolation is expected to overestimate the true 150~GHz flux density for most AGN \citep{Hogan15b}. Our goal is therefore to quantify this overestimation. We assume that uncertainties in the X-ray-derived model of the SZ signal are subdominant compared to the systematic error introduced by the power-law extrapolation of the AGN emission. We subtract the best-fit SZ model from the observed ACT maps at the cluster locations, leaving residual emission dominated by the AGN. We then fit a Gaussian profile to this residual and adopt the resulting best-fit flux density as our measurement of the 150 GHz AGN emission.
For MACS0242, the flux density obtained after cluster subtraction is $55.8 \pm 6.8$ mJy, fully consistent with the extrapolated value. For RXJ0439, the recovered flux density is $64.0 \pm 7.6$ mJy, corresponding to only a $\approx 4\%$ underestimate relative to the extrapolation. For both of the clusters, we estimate the statistical noise in the flux density measurement by fitting a Gaussian profile to random noise patches within $20^\circ$ of the cluster center in the ACT map. Finally, subtracting both the measured AGN flux density and the SZ signal yields residual ACT maps consistent with noise. Across these two AGN systems, which exhibit different SEDs, we therefore find little to no evidence for systematic overestimation in the 150 GHz flux density inferred from extrapolations based on 30 and 90 GHz measurements.
\\\\
\subsection{Millimeter SED fits to the AGN}
\begin{figure*}
    \begin{subfigure}[t]{0.45\linewidth}
        \centering
        \includegraphics[width=\linewidth]{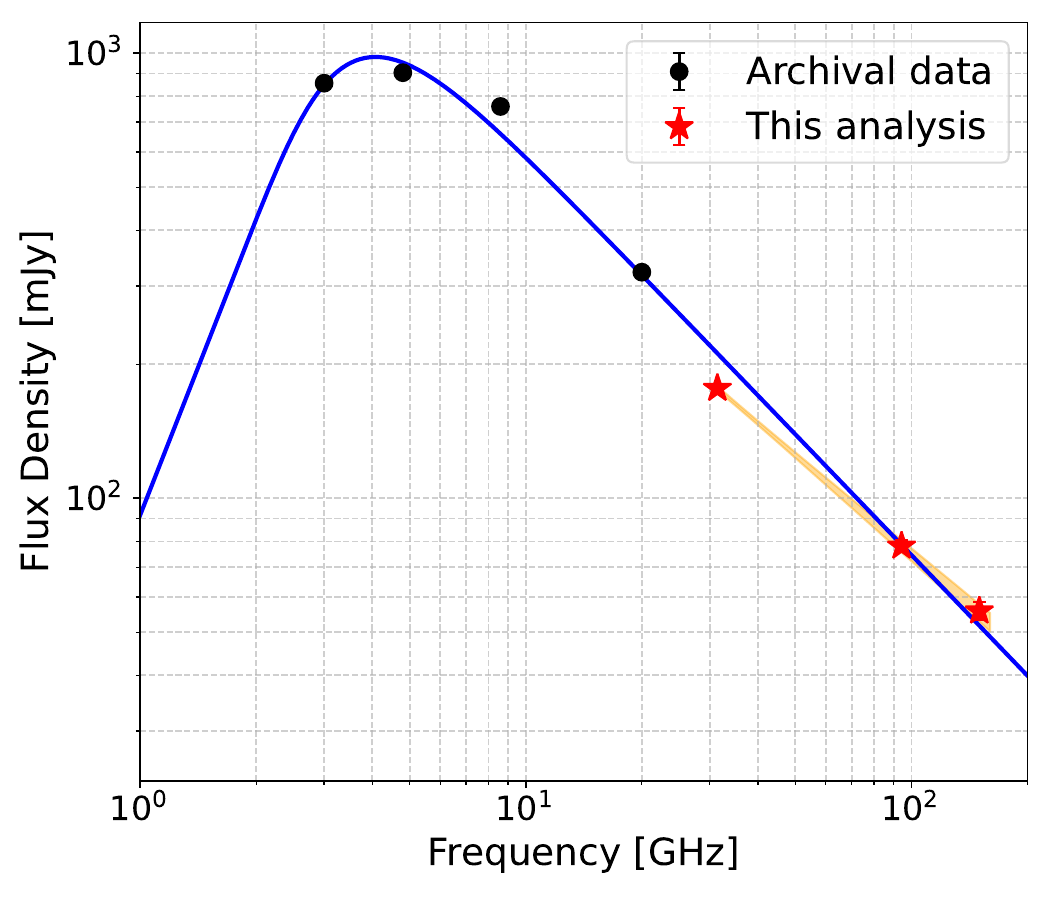}
        \caption{MACS0242 SED fits. Archival data is from \cite{VLASS, AT20G, Hogan15b}}
    \label{fig:MACS0242_SED}
    \end{subfigure}
    \hfill
    \begin{subfigure}[t]{0.45\linewidth}
        \centering
        \includegraphics[width=\linewidth]{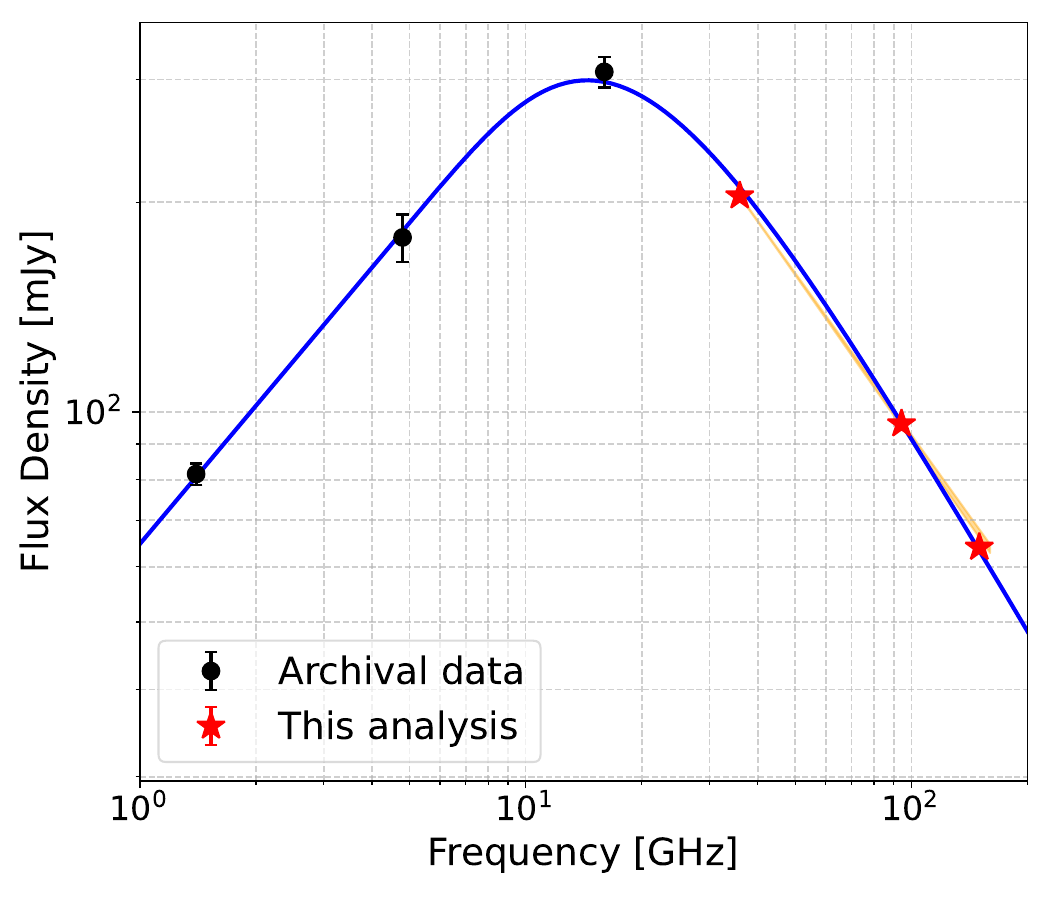}
        \caption{RXJ0439 SED fits. Archival data is from \cite{NVSS, PMN, CRATES, Hogan15b} }
    \label{fig:RXJ0439_SED}
    \end{subfigure}
    \caption{SED fits in blue to the flux densities of the AGN cores in the CGs of MACS0242 and RXJ0439 using Eqn. \ref{eqn:SED}, with black points corresponding to previous measurements from the literature, and red stars corresponding to our measurements, including those obtained at 150~GHz from the ACT data based on subtracting an X-ray-derived model of the SZ signal. We note that for some archival points we do not have measurement errors, so we assume a fractional uncertainty of $10\%$ for all data points for the fit. The shaded orange region corresponds to the $68\%$ confidence interval for the simple linear extrapolation from our CARMA measurements. }
\end{figure*}

We utilize our flux density measurements of the central AGN cores, along with archival observations at GHz frequencies, to model their radio spectral energy distributions (SEDs) using a physically motivated prescription combining a double power-law–like synchrotron spectrum with internal free–free absorption. This approach is commonly utilized for radio AGN when the observed data span the peak in emission \citep{AGNfitter,SED_fitting}. The model spectrum includes a power law component described by a slope, $\alpha$, modulated by free-free absorption in a thermal plasma that coexists with the relativistic electrons responsible for the synchrotron emission. This is described via an escape probability formalism, using an opacity, $\tau_\nu$, dominated by free-free emission. Thus, the spectral flux density ($S_\nu$) can be expressed as
\begin{align}\label{eqn:SED}
    S_\nu &= A_{10} \bigg(\frac{\nu}{10}\bigg)^{\alpha} \bigg( \frac{1 - e^{-\tau_\nu}}{\tau_\nu}\bigg); \\
    \tau_\nu &= \bigg(\frac{\nu}{\nu_0}\bigg)^{-\beta}.
\end{align}
The 4 parameters in this model are: $A_{10}$ - the normalization constant corresponding to the amplitude at 10 GHz, $\alpha$ - the power exponent, $\nu_0$ - the turnover frequency at which the source becomes optically thick, and $\beta$ - the power exponent for the opacity. We use only 1--150\,GHz data in order to avoid potential contamination from extended emission, which will be brighter and less well resolved at lower frequencies. The resulting SED fits for the AGN in the CGs of MACS0242 and RXJ0439 are given in Fig \ref{fig:MACS0242_SED} and \ref{fig:RXJ0439_SED} respectively. We note that the SED shapes for these two AGN are very different, and, for example, any attempt to extrapolate typical low-frequency radio observations at $\lesssim 5$~GHz to millimeter wavelengths near 100~GHz based on assuming a single SED shape would in general be extremely inaccurate, however we note that a simple power law extrapolation above $\sim 20$ GHz provides lower extrapolation errors.

\subsection{Temporal variability of the AGN Cores}
\begin{figure}
    \centering
    \includegraphics[width=\linewidth]{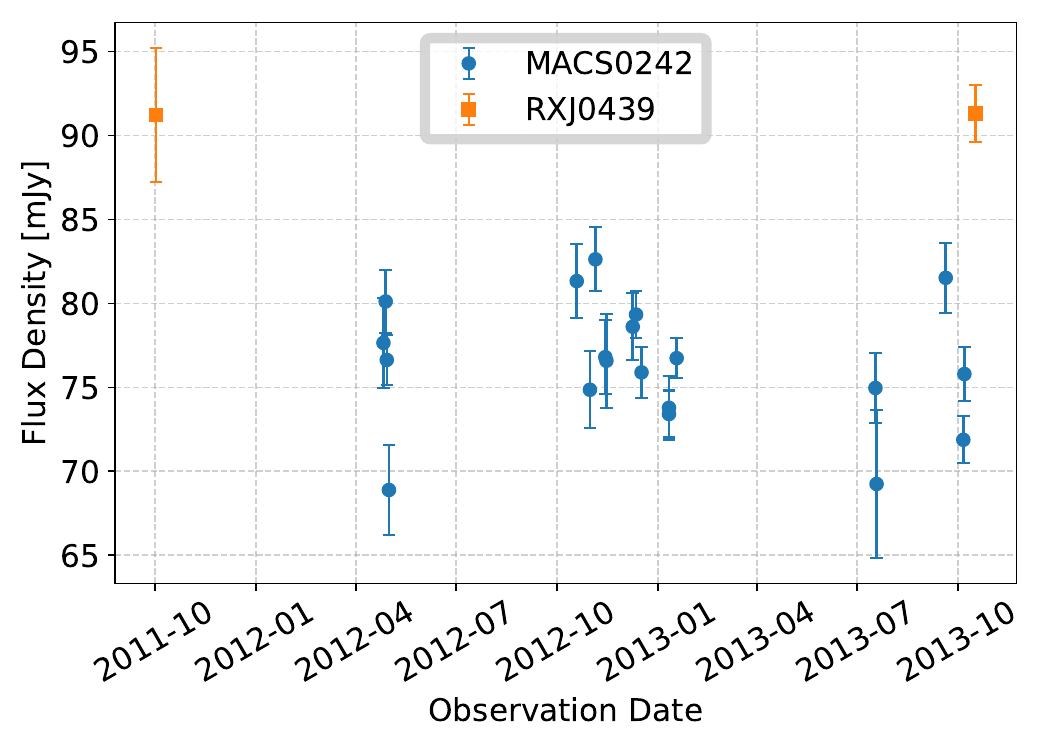}
    \caption{Lightcurves of the AGN cores within the CGs of MACS0242 and RXJ0439, spanning our observations from April 2012 - October 2013 and October 2011-October2013 respectively.}
    \label{fig:lightcurve}
\end{figure}

Because our 90 GHz CARMA observations span years, we examine the temporal variability of the radio emission from the AGN cores in both clusters. Short-timescale variability places constraints on the physical size of the emitting region, while longer-term (multi-year) variations provide insight into the fueling and duty cycle of AGN activity \citep{Hogan15b}. 
The 90 GHz light curves of both MACS0242 and RXJ0439 are shown in Fig. \ref{fig:lightcurve}.
\\\\
The light curve of MACS0242 exhibits a maximum fractional variability \citep{frac_var} of approximately 4\% over the full duration of the CARMA observations, spanning the period April 2012 to October 2013. This level of variability is significantly smaller than that inferred from GISMO measurements reported by \cite{Hogan15b}, who find a $\sim$50\% decrease in the 150 GHz flux density between April 2012 and April 2013 based on a single measurement at each of those epochs. Our CARMA data span April–December 2012 and July–October 2013, over which we observe only modest variability at 90 GHz. We note that we do not have coverage in April 2013; therefore we cannot rule out their measurement, but the apparent flux density drop noted by \cite{Hogan15b} may be associated with observational systematics. In particular, \cite{Hogan15b} report an unusually high value for the millimeter-wave optical depth of the atmosphere during their April 2013 observations. For RXJ0439, our temporal coverage is limited to two epochs, in October 2011 and October 2013, as shown in Fig. \ref{fig:lightcurve}. The measured flux densities at these epochs are consistent within their uncertainties, and so there is no statistically significant evidence for variability. This behavior is in agreement with the findings of \cite{Hogan15b}, who similarly report a lack of strong variability for this source between April 2012 to April 2013 at 150 GHz. 

\subsection{The AGN in RXJ1651 (Hercules A)}
\begin{figure}
    \centering
    \includegraphics[width=\linewidth]{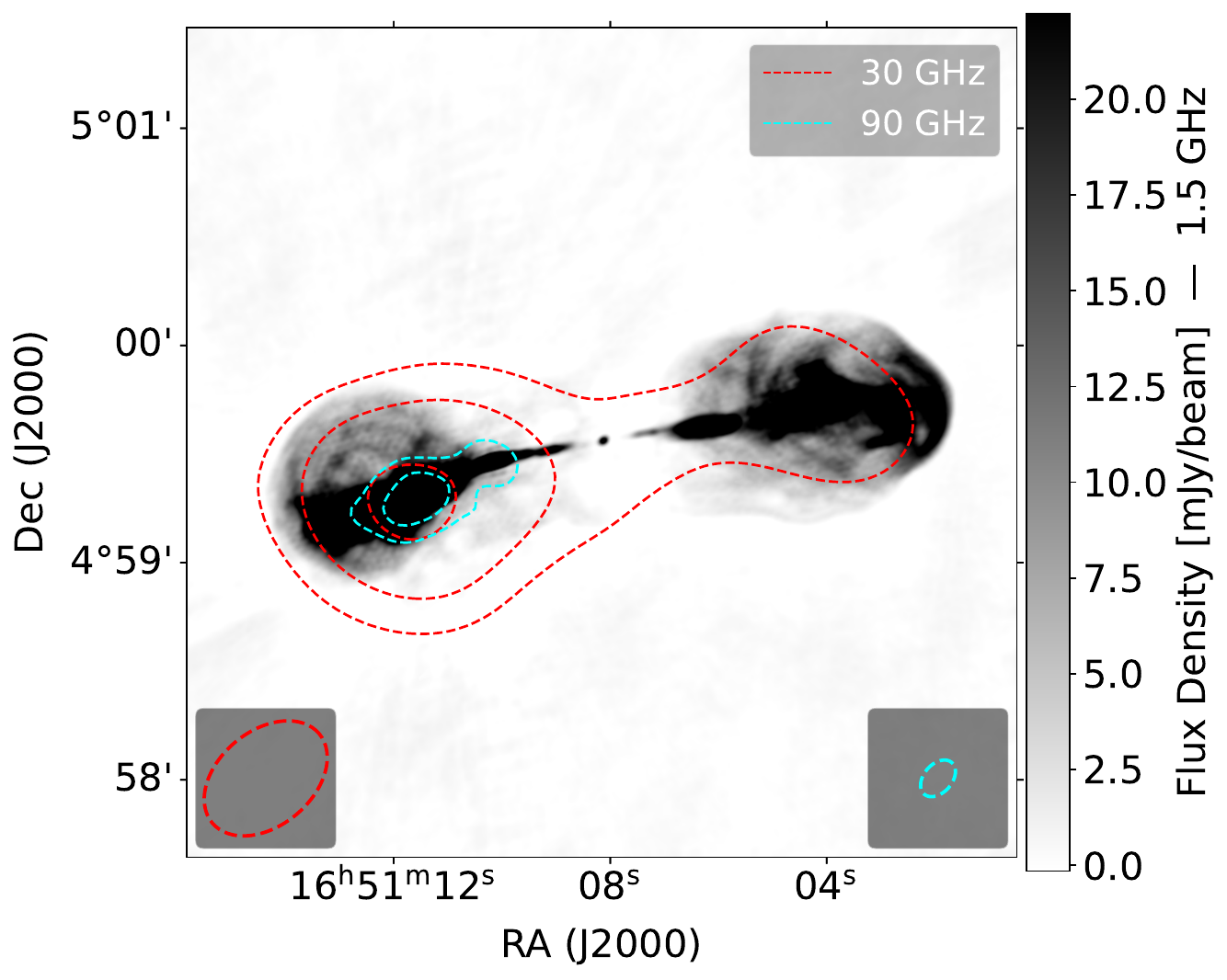}
    \caption{1.5 GHz emission from Hercules A, at the center of RXJ1615, as observed by \cite{Timmerman_2023}. We overlay $10\sigma$, $30\sigma$ and $100\sigma$ ($30$ GHz only) contours from our CLEAN maps of the CARMA data at 30 and 90 GHz in red and cyan respectively. We also add elliptical dashed patches showing the synthesized beam at 30 and 90 GHz in red and cyan respectively in the bottom corners.}
    \label{fig:HercA_plot}
\end{figure}

For the AGN core of Hercules A, unlike our other targets, both CARMA bands reveal extended emission associated with the radio lobes.
Consequently, we cannot fit a simple point-source model to the visibilities, as we do for the other clusters.
Instead, we use the CLEAN algorithm \citep{CLEAN_radio} to produce a model for the observed emission, adopting a gain parameter of 0.05 and terminating the procedure when the amplitude of the next CLEAN component would be negative.
At 30 GHz there is structure clearly detected even at the shortest baselines (largest scales), and so we included all baselines in the fit rather than restricting to $uv$ radii $>2$k$\lambda$, as we do elsewhere.
Contours of the CLEAN maps are overlaid on a 1.5\,GHz image in Figure~\ref{fig:HercA_plot}.

Neither CLEAN model incorporates emission from the nucleus itself.
The morphology of the 30\,GHz emission broadly follows the lobes detected at lower frequencies, while the higher-resolution 90\,GHz model appears to be associated with a bright feature near the interface between the eastern jet and the corresponding radio lobe \citep{Timmerman_2023}.
\\\\
Like the central AGN cores in MACS0242 and RXJ0439, the AGN emission from the eastern lobe of Hercules A is also much larger than the magnitude of the SZ signal at 90\,GHz, although in the case of Hercules A the emission is much more extended with a complicated morphology. However, given the angular resolution of ACT, the lobe emission is effectively unresolved, and we therefore model it as a point-like source for our subsequent analysis. For Hercules A, we initially estimate the total flux density directly from the CLEAN maps. We degrade the angular resolution of the 90 GHz CLEAN map to the angular resolution of the 30 GHz map, and perform aperture photometry using a user-defined elliptical aperture slightly larger than the degraded synthesized beam at 90 GHz, chosen to maximize the recovered flux density from the bright knot in the eastern lobe. This yields a flux density estimate of 293.3 mJy at 94 GHz, and 1024.4 mJy at 30 GHz within the corresponding aperture. Subtracting a point-like model based on a power-law extrapolation of these flux density measurements from the 98 GHz ACT map still results in a negative cluster SNR, indicating that the CLEAN-based flux density estimate at 90 GHz is underestimated. Some underestimation is expected, however, as there could well be even more extended 90 GHz emission resolved out of the CARMA data in this system.
\\\\
To obtain a more accurate total flux density measurement, we instead follow the approach described in Section \ref{150_overest} to measure 150~GHz flux densities for the AGN cores in MACS0242 and RXJ0439, this time applying it to the 98~GHz ACT map. Specifically, we subtract an X-ray-derived model of the SZ signal from the 98~GHz ACT map and fit a Gaussian beam profile to obtain the best-fit flux density at the location of Hercules A. Applying this correction yields a final flux density of $310.6 \pm 9.7$ mJy in the ACT 98 GHz band. We repeat the same procedure for the ACT 150 GHz band and obtain a corrected flux density of $163.8 \pm 13.7$ mJy. Unlike the previous two sources, we note clear residuals in the cluster and AGN subtracted map for both 98 and 150 GHz for Hercules A, possibly due to the complex and extended morphology of the source. Given the presence of multiple components with both steep and shallow spectral indices, we do not attempt to fit a single SED model to Hercules A. Nevertheless, our corrected high-frequency flux density measurements are consistent with Planck 90 GHz observations of Hercules A, which provide the only other constraint on this source at comparable frequencies \citep{Planck_100_sources}.

\section{Conclusion}\label{conc}
In this work, we present measurements of the 30 and 90 GHz flux densities of bright AGN in the CGs in three galaxy clusters—MACS0242, RXJ0439, and RXJ1615—using CARMA. None of these clusters are detected from a standard SZ analysis of ACT survey data based on the MMF3 detection algorithm, but the first two clusters are detected with high SNR after subtraction of the AGN emission. We further derive flux density estimates for the AGN cores at 150 GHz by subtracting the cluster SZ signal from the ACT maps based on X-ray-derived models, thereby extending the SEDs of these AGN cores into the millimeter regime. In addition, we investigate temporal variability using time-asynchronous observations for two of the sources lacking complex, extended millimeter emission.
\\\\
We present a methodology for estimating millimeter flux densities of radio-bright AGNs in CGs that leverages independent constraints on the cluster SZ signal. Using X-ray–derived cluster models, we subtract the SZ contribution from ACT maps and measure the residual AGN emission by modeling it as a point source. This approach provides robust flux density estimates for bright AGNs that dominate over the underlying SZ signal, although its applicability is more limited for fainter sources that only partially contaminate the cluster signal. In the absence of direct millimeter measurements, this method reduces systematic uncertainties relative to simple SED extrapolation by anchoring the analysis to well-constrained X-ray cluster properties. Repeating this procedure for a larger statistical sample would enable us to better calibrate the correction factor between true millimeter flux and extrapolated flux from lower frequencies, thereby resulting in improved extrapolation estimates at millimeter frequencies relevant for SZ surveys.
\\\\
Despite the critical role of AGN feedback in cluster evolution, few observational studies have focused on the millimeter-wave ($\sim30$–300 GHz) emission of the AGN hosted in CGs \citep{Hogan15b}. We observe different SED shapes, demonstrating that even within our small sample, the inferred flux densities at higher frequencies would be mis-estimated from simple power-law extrapolations from lower ($\sim$ GHz) frequencies. We also detect fractional variability at the $\sim4\%$ level in one source, introducing an additional source of uncertainty when non-contemporaneous data are used. Together, these effects underscore the necessity of contemporaneous, high-resolution, and high frequency millimeter observations for accurate accounting of AGN contamination in SZ surveys. In future papers in this series, we will build on the methodology and results presented here, together with forthcoming datasets, to quantitatively assess the impact of AGN contamination on SZ-selected cluster samples.
\\\\
This issue is expected to worsen at higher redshifts, as the fraction of AGN-hosting CGs increases over $0<z<1.3$ \citep{Somboonpanyakul_2022}. Estimates by \cite{Hogan15b} suggest that $\gtrsim6\%$ of all clusters host millimeter-bright AGNs, rising to $\sim15\%$ for cool-core systems. Thus, accurately quantifying AGN contamination is essential for precision cluster cosmology. This need is particularly pressing for upcoming deep SZ surveys such as SPT 3G \citep{SPT3G} and the Simons Observatory \citep{SO}, which aim to deliver high redshift cluster samples for cosmology.

\begin{acknowledgments}
The authors are grateful to Tom Crawford and Carolyn Topper for their assistance in finding and recovering the CARMA data from backup. HS and JS were partially supported by NASA award 80NSSC25K0597. DY was supported by the Leadership Alliance Summer Research Early Identification program. We acknowledge support from the U.S.\ Department of Energy under contract number DE-AC02-76SF00515.
\end{acknowledgments}

\facility{CARMA}
\software{Astropy \citep{The_Astropy_Collaboration_2022}, Difmap \citep{difmap}, Matplotlib \citep{matplotlib}, Scipy \citep{scipy}}

\appendix

\section{Target List}\label{targets_list}
\begin{table*}[ht]
    \centering
    \caption{List of targets with details of the archival CARMA data}
    \label{tab:data}
    \begin{tabular}{|c|c|c|c|c|c|}
        \hline
        \textbf{Cluster} & \textbf{Track ID} & \textbf{Obs. date (mm/dd/yy)} & \textbf{On time (mins)} & \textbf{Freq. (GHz)} & \textbf{Flux cal.}\\
        \hline
        RXJ0439 & c0769Z.4SH\_30RXJ043.2 & 08/23/11 & 75 & 34.9 & Mars\\
        \hline
        RXJ0439 & c0769Z.4SH\_30RXJ043.3 & 08/24/11 & 200 & 34.9 & Mars \\ 
        \hline
        RXJ0439 & c0769Z.4SH\_30RXJ043.4 & 08/31/11 & 420 & 34.9 & Uranus \\ 
        \hline
        RXJ0439 & c0769Z.5SH\_90RXJ043.3 & 10/02/11 & 100 & 93.0 & Mars \\ 
        \hline
        RXJ0439 & c1173.16SH\_90r04390.5 & 10/17/13 & 420 & 88.0 & Uranus \\ 
        \hline
        MACS0242 & c0950.4SL\_31J02422.1 & 04/12/12 & 230 & 34.9 & Neptune \\ 
        \hline
        MACS0242 & c0950.4SL\_90J02422.2 & 04/28/12 & 210 & 90.8 & Uranus \\ 
        \hline
        MACS0242 & c0950.4SL\_90J02422.3 & 04/29/12 & 300 & 90.8 & Uranus \\ 
        \hline
        MACS0242 & c0950.4SL\_90J02422.4 & 05/01/12 & 240 & 90.8 & Uranus \\ 
        \hline
        MACS0242 & c1009.3SL\_90MACSJ0.1 & 10/19/12 & 270 & 86.5 & Uranus \\ 
        \hline
        MACS0242 & c1009.3SL\_90MACSJ0.2 & 10/31/12 & 360 & 86.5 & Uranus \\ 
        \hline
        MACS0242 & c1009.3SL\_90MACSJ0.3 & 11/05/12 & 300 & 86.5 & Uranus \\ 
        \hline
        MACS0242 & c1009.3SL\_90MACSJ0.4 & 11/14/12 & 300 & 86.5 & Uranus \\ 
        \hline
        MACS0242 & c1009.3SL\_90MACSJ0.5 & 11/15/12 & 240 & 86.5 & Uranus \\ 
        \hline
        MACS0242 & c1009.3SL\_90MACSJ0.6 & 12/09/12 & 240 & 86.5 & Uranus \\ 
        \hline
        MACS0242 & c1009.3SL\_90MACSJ0.7 & 12/12/12 & 270 & 86.5 & Uranus \\ 
        \hline
        MACS0242 & c1009.3SL\_90MACSJ0.8 & 12/17/12 & 360 & 86.5 & Uranus \\ 
        \hline
        MACS0242 & c1009.3SL\_90MACSJ0.9 & 01/11/13 & 240 & 86.5 & Uranus \\ 
        \hline
        MACS0242 & c1009.3SL\_90MACSJ0.10 & 01/11/13 & 300 & 86.5 & Uranus \\ 
        \hline
        MACS0242 & c1009.3SL\_90MACSJ0.11 & 01/18/13 & 270 & 86.5 & Uranus \\ 
        \hline
        MACS0242 & c1098.12SL\_90MACS02.1 & 07/18/13 & 270 & 86.5 & Uranus \\ 
        \hline
        MACS0242 & c1098.12SL\_90MACS02.8  & 09/20/13 & 300 & 86.5 & Uranus \\ 
        \hline
        MACS0242 & c1098.12SL\_90MACS02.10 & 10/07/13 & 300 & 86.5 & Uranus \\ 
        \hline
        MACS0242 & c1098.12SL\_90MACS02.9  & 10/06/13 & 270 & 86.5 & Uranus \\ 
        \hline
        MACS0242 & c1098.12SL\_90MACS02.2 & 10/19/13 & 270 & 86.5 & Uranus \\ 
        \hline
        RXJ1615 & c1009.46SL\_30RXJ165.1 & 09/19/12 & 240 & 34.9 & Mars \\ 
        \hline
        RXJ1615 & c1009.46SL\_30RXJ165.3 & 09/23/13 & 240 & 34.9 & Mars \\ 
        \hline
        RXJ1615 & c1009.46SL\_30RXJ165.4 & 09/24/13 & 240 & 34.9 & Mars \\ 
        \hline
        RXJ1615 & c1009.46SL\_30RXJ165.6 & 09/27/13 & 150 & 34.9 & Mars \\ 
        \hline
        RXJ1615 & c1009.46SL\_30RXJ165.7 & 09/28/13 & 75 & 34.9 & Neptune \\ 
        \hline
        RXJ1615 & c1173.3SH\_90r16510.11 & 10/26/13 & 210 & 89.0 & Neptune \\ 
        \hline
        RXJ1615 & c1173.3SH\_90r16510.14 & 10/27/13 & 165 & 89.0 & Mars \\ 
        \hline
        RXJ1615 & c1173.3SH\_90r16510.15 & 10/28/13 & 240 & 89.0 & Neptune \\ 
        \hline
        RXJ1615 & c1173.3SH\_90r16510.3 & 10/13/13 & 225 & 88.0 & Neptune \\ 
        \hline
        RXJ1615 & c1173.3SH\_90r16510.4 & 10/14/13 & 225 & 88.0 & Neptune \\ 
        \hline
        RXJ1615 & c1173.3SH\_90r16510.5 & 10/15/13 & 6 & 88.0 & Mars \\ 
        \hline
        RXJ1615 & c1173.3SH\_90r16510.7 & 10/17/13 & 360 & 89.0 & Mars \\ 
        \hline
        RXJ1615 & c1173.3SH\_90r16510.8 & 10/21/13 & 240 & 89.0 & Neptune \\ 
        \hline
        RXJ1615 & c1173.3SH\_90r16510.9 & 10/23/13 & 240 & 89.0 & Neptune \\ 
        \hline
    \end{tabular}
\end{table*}

\bibliography{sample7}{}

@article{Hardcastle0804.3369,
	adsurl = {https://ui.adsabs.harvard.edu/abs/2008MNRAS.388..176H},
	archiveprefix = {arXiv},
	author = {{Hardcastle}, M.~J. and {Looney}, L.~W.},
	doi = {10.1111/j.1365-2966.2008.13370.x},
	eprint = {0804.3369},
	journal = {\mnras},
	month = jul,
	number = {1},
	pages = {176-186},
	primaryclass = {astro-ph},
	title = {{The properties of powerful radio sources at 90 GHz}},
	volume = {388},
	year = 2008}

@article{Zhuravleva1707.02304,
	adsurl = {https://ui.adsabs.harvard.edu/abs/2018ApJ...865...53Z},
	archiveprefix = {arXiv},
	author = {{Zhuravleva}, I. and {Allen}, S.~W. and {Mantz}, A. and {Werner}, N.},
	doi = {10.3847/1538-4357/aadae3},
	eid = {53},
	eprint = {1707.02304},
	journal = {\apj},
	month = sep,
	pages = {53},
	primaryclass = {astro-ph.HE},
	title = {{Gas Perturbations in the Cool Cores of Galaxy Clusters: Effective Equation of State, Velocity Power Spectra, and Turbulent Heating}},
	volume = 865,
	year = 2018}

@article{Zhuravleva1410.6485,
	adsurl = {http://adsabs.harvard.edu/abs/2014Natur.515...85Z},
	archiveprefix = {arXiv},
	author = {{Zhuravleva}, I. and {Churazov}, E. and {Schekochihin}, A.~A. and {Allen}, S.~W. and {Ar{\'e}valo}, P. and {Fabian}, A.~C. and {Forman}, W.~R. and {Sanders}, J.~S. and {Simionescu}, A. and {Sunyaev}, R. and {Vikhlinin}, A. and {Werner}, N.},
	doi = {10.1038/nature13830},
	eprint = {1410.6485},
	journal = {\nat},
	month = nov,
	pages = {85-87},
	primaryclass = {astro-ph.HE},
	title = {{Turbulent heating in galaxy clusters brightest in X-rays}},
	volume = 515,
	year = 2014}

@article{Rykoff1303.3562,
	adsurl = {http://adsabs.harvard.edu/abs/2014ApJ...785..104R},
	archiveprefix = {arXiv},
	author = {{Rykoff}, E.~S. and {Rozo}, E. and {Busha}, M.~T. and {Cunha}, C.~E. and {Finoguenov}, A. and {Evrard}, A. and {Hao}, J. and {Koester}, B.~P. and {Leauthaud}, A. and {Nord}, B. and {Pierre}, M. and {Reddick}, R. and {Sadibekova}, T. and {Sheldon}, E.~S. and {Wechsler}, R.~H.},
	doi = {10.1088/0004-637X/785/2/104},
	eid = {104},
	eprint = {1303.3562},
	journal = {\apj},
	month = apr,
	pages = {104},
	title = {{redMaPPer. I. Algorithm and SDSS DR8 Catalog}},
	volume = 785,
	year = 2014}

@article{Rykoff1601.00621,
	adsurl = {http://adsabs.harvard.edu/abs/2016ApJS..224....1R},
	archiveprefix = {arXiv},
	author = {{Rykoff}, E.~S. and {Rozo}, E. and {Hollowood}, D. and {Bermeo-Hernandez}, A. and {Jeltema}, T. and {Mayers}, J. and {Romer}, A.~K. and {Rooney}, P. and {Saro}, A. and {Vergara Cervantes}, C. and {Wechsler}, R.~H. and {Wilcox}, H. and {Abbott}, T.~M.~C. and {Abdalla}, F.~B. and {Allam}, S. and {Annis}, J. and {Benoit-L{\'e}vy}, A. and {Bernstein}, G.~M. and {Bertin}, E. and {Brooks}, D. and {Burke}, D.~L. and {Capozzi}, D. and {Carnero Rosell}, A. and {Carrasco Kind}, M. and {Castander}, F.~J. and {Childress}, M. and {Collins}, C.~A. and {Cunha}, C.~E. and {D'Andrea}, C.~B. and {da Costa}, L.~N. and {Davis}, T.~M. and {Desai}, S. and {Diehl}, H.~T. and {Dietrich}, J.~P. and {Doel}, P. and {Evrard}, A.~E. and {Finley}, D.~A. and {Flaugher}, B. and {Fosalba}, P. and {Frieman}, J. and {Glazebrook}, K. and {Goldstein}, D.~A. and {Gruen}, D. and {Gruendl}, R.~A. and {Gutierrez}, G. and {Hilton}, M. and {Honscheid}, K. and {Hoyle}, B. and {James}, D.~J. and {Kay}, S.~T. and {Kuehn}, K. and {Kuropatkin}, N. and {Lahav}, O. and {Lewis}, G.~F. and {Lidman}, C. and {Lima}, M. and {Maia}, M.~A.~G. and {Mann}, R.~G. and {Marshall}, J.~L. and {Martini}, P. and {Melchior}, P. and {Miller}, C.~J. and {Miquel}, R. and {Mohr}, J.~J. and {Nichol}, R.~C. and {Nord}, B. and {Ogando}, R. and {Plazas}, A.~A. and {Reil}, K. and {Sahl{\'e}n}, M. and {Sanchez}, E. and {Santiago}, B. and {Scarpine}, V. and {Schubnell}, M. and {Sevilla-Noarbe}, I. and {Smith}, R.~C. and {Soares-Santos}, M. and {Sobreira}, F. and {Stott}, J.~P. and {Suchyta}, E. and {Swanson}, M.~E.~C. and {Tarle}, G. and {Thomas}, D. and {Tucker}, D. and {Uddin}, S. and {Viana}, P.~T.~P. and {Vikram}, V. and {Walker}, A.~R. and {Zhang}, Y. and {DES Collaboration}},
	doi = {10.3847/0067-0049/224/1/1},
	eid = {1},
	eprint = {1601.00621},
	journal = {\apjs},
	month = may,
	pages = {1},
	title = {{The RedMaPPer Galaxy Cluster Catalog From DES Science Verification Data}},
	volume = 224,
	year = 2016}

@article{Planck1502.01597,
	adsurl = {http://adsabs.harvard.edu/abs/2016A%26A...594A..24P},
	archiveprefix = {arXiv},
	author = {{Planck Collaboration} and {Ade}, P.~A.~R. and {Aghanim}, N. and {Arnaud}, M. and {Ashdown}, M. and {Aumont}, J. and {Baccigalupi}, C. and {Banday}, A.~J. and {Barreiro}, R.~B. and {Bartlett}, J.~G. and et al.},
	doi = {10.1051/0004-6361/201525833},
	eid = {A24},
	eprint = {1502.01597},
	journal = {\aap},
	month = sep,
	pages = {A24},
	title = {{Planck 2015 results. XXIV. Cosmology from Sunyaev-Zeldovich cluster counts}},
	volume = 594,
	year = 2016}

@article{Mantz1407.4516,
	adsurl = {http://adsabs.harvard.edu/abs/2015MNRAS.446.2205M},
	archiveprefix = {arXiv},
	author = {{Mantz}, A.~B. and {von der Linden}, A. and {Allen}, S.~W. and {Applegate}, D.~E. and {Kelly}, P.~L. and {Morris}, R.~G. and {Rapetti}, D.~A. and {Schmidt}, R.~W. and {Adhikari}, S. and {Allen}, M.~T. and {Burchat}, P.~R. and {Burke}, D.~L. and {Cataneo}, M. and {Donovan}, D. and {Ebeling}, H. and {Shandera}, S. and {Wright}, A.},
	doi = {10.1093/mnras/stu2096},
	eprint = {1407.4516},
	journal = {\mnras},
	month = jan,
	pages = {2205-2225},
	title = {{Weighing the giants - IV. Cosmology and neutrino mass}},
	volume = 446,
	year = 2015}

@ARTICLE{Allen2011,
       author = {{Allen}, Steven W. and {Evrard}, August E. and {Mantz}, Adam B.},
        title = "{Cosmological Parameters from Observations of Galaxy Clusters}",
      journal = {\araa},
         year = 2011,
        month = sep,
       volume = {49},
       number = {1},
        pages = {409-470},
          doi = {10.1146/annurev-astro-081710-102514},
archivePrefix = {arXiv},
       eprint = {1103.4829},
 primaryClass = {astro-ph.CO},
       adsurl = {https://ui.adsabs.harvard.edu/abs/2011ARA\&A..49..409A}
}

@article{Lesci_2022,
   title={AMICO galaxy clusters in KiDS-DR3: Cosmological constraints from counts and stacked weak lensing},
   volume={659},
   ISSN={1432-0746},
   url={http://dx.doi.org/10.1051/0004-6361/202040194},
   DOI={10.1051/0004-6361/202040194},
   journal={Astronomy \&amp; Astrophysics},
   publisher={EDP Sciences},
   author={Lesci, G. F. and Marulli, F. and Moscardini, L. and Sereno, M. and Veropalumbo, A. and Maturi, M. and Giocoli, C. and Radovich, M. and Bellagamba, F. and Roncarelli, M. and Bardelli, S. and Contarini, S. and Covone, G. and Ingoglia, L. and Nanni, L. and Puddu, E.},
   year={2022},
   month=mar, pages={A88} }

@article{Ghirardini_2024,
   title={The SRG/eROSITA all-sky survey: Cosmology constraints from cluster abundances in the western Galactic hemisphere},
   volume={689},
   ISSN={1432-0746},
   url={http://dx.doi.org/10.1051/0004-6361/202348852},
   DOI={10.1051/0004-6361/202348852},
   journal={Astronomy \& Astrophysics},
   publisher={EDP Sciences},
   author={Ghirardini, V. and Bulbul, E. and Artis, E. and Clerc, N. and Garrel, C. and Grandis, S. and Kluge, M. and Liu, A. and Bahar, Y. E. and Balzer, F. and Chiu, I. and Comparat, J. and Gruen, D. and Kleinebreil, F. and Krippendorf, S. and Merloni, A. and Nandra, K. and Okabe, N. and Pacaud, F. and Predehl, P. and Ramos-Ceja, M. E. and Reiprich, T. H. and Sanders, J. S. and Schrabback, T. and Seppi, R. and Zelmer, S. and Zhang, X. and Bornemann, W. and Brunner, H. and Burwitz, V. and Coutinho, D. and Dennerl, K. and Freyberg, M. and Friedrich, S. and Gaida, R. and Gueguen, A. and Haberl, F. and Kink, W. and Lamer, G. and Li, X. and Liu, T. and Maitra, C. and Meidinger, N. and Mueller, S. and Miyatake, H. and Miyazaki, S. and Robrade, J. and Schwope, A. and Stewart, I.},
   year={2024},
   month=sep, pages={A298} }

@ARTICLE{DESY3_2025_Halo_Abundance,
       author = {{DES Collaboration} and {Abbott}, T.~M.~C. and {Aguena}, M. and {Alarcon}, A. and {Anbajagane}, D. and {Andrade-Oliveira}, F. and {Avila}, S. and {Bacon}, D. and {Becker}, M.~R. and {Bhargava}, S. and {Blazek}, J. and {Bocquet}, S. and {Brooks}, D. and {Carnero Rosell}, A. and {Carretero}, J. and {Castander}, F.~J. and {Chang}, C. and {Choi}, A. and {Conselice}, C. and {Costanzi}, M. and {Crocce}, M. and {da Costa}, L.~N. and {Pereira}, M.~E.~S. and {Davis}, T.~M. and {Desai}, S. and {Diehl}, H.~T. and {Dodelson}, S. and {Doel}, P. and {Elvin-Poole}, J. and {Esteves}, J. and {Everett}, S. and {Farahi}, A. and {Fert{\'e}}, A. and {Flaugher}, B. and {Garc{\'\i}a-Bellido}, J. and {Gatti}, M. and {Giannini}, G. and {Giles}, P. and {Grandis}, S. and {Gruen}, D. and {Gruendl}, R.~A. and {Gutierrez}, G. and {Harrison}, I. and {Hinton}, S.~R. and {Hollowood}, D.~L. and {Honscheid}, K. and {Jeffrey}, N. and {Jeltema}, T. and {Krause}, E. and {Lahav}, O. and {Lee}, S. and {Lidman}, C. and {Lima}, M. and {Lin}, H. and {Mohr}, J.~J. and {Marshall}, J.~L. and {McCullough}, J. and {Mena-Fern}, J. and {Miquel}, R. and {Muir}, J. and {Myles}, J. and {Ogando}, R.~L.~C. and {Palmese}, A. and {Paterno}, M. and {Plazas Malag{\'o}n}, A.~A. and {Porredon}, A. and {Prat}, J. and {Romer}, A.~K. and {Roodman}, A. and {Rozo}, E. and {Rykoff}, E.~S. and {Sanchez}, E. and {Sanchez Cid}, D. and {Sevilla-Noarbe}, I. and {Smith}, M. and {Suchyta}, E. and {Tarle}, G. and {Thomas}, D. and {To}, Chun-Hao and {Troxel}, M.~A. and {Vikram}, V. and {Walker}, A.~R. and {Weinberg}, David H. and {Weaverdyck}, N. and {Wechsler}, R.~H. and {Weller}, J. and {Wu}, H. -Y. and {Yamamoto}, M. and {Yanny}, B. and {Zhang}, Y. and {Zhou}, C.},
        title = "{Dark Energy Survey Year 3 Results: Cosmological Constraints from Cluster Abundances, Weak Lensing, and Galaxy Clustering}",
      journal = {arXiv e-prints},
         year = 2025,
        month = mar,
          eid = {arXiv:2503.13632},
        pages = {arXiv:2503.13632},
          doi = {10.48550/arXiv.2503.13632},
archivePrefix = {arXiv},
       eprint = {2503.13632},
 primaryClass = {astro-ph.CO},
       adsurl = {https://ui.adsabs.harvard.edu/abs/2025arXiv250313632D}
}

@ARTICLE{Bocquet2024,
       author = {{Bocquet}, S. and {Grandis}, S. and {Bleem}, L.~E. and {Klein}, M. and {Mohr}, J.~J. and {Schrabback}, T. and {Abbott}, T.~M.~C. and {Ade}, P.~A.~R. and {Aguena}, M. and {Alarcon}, A. and {Allam}, S. and {Allen}, S.~W. and {Alves}, O. and {Amon}, A. and {Anderson}, A.~J. and {Annis}, J. and {Ansarinejad}, B. and {Austermann}, J.~E. and {Avila}, S. and {Bacon}, D. and {Bayliss}, M. and {Beall}, J.~A. and {Bechtol}, K. and {Becker}, M.~R. and {Bender}, A.~N. and {Benson}, B.~A. and {Bernstein}, G.~M. and {Bhargava}, S. and {Bianchini}, F. and {Brodwin}, M. and {Brooks}, D. and {Bryant}, L. and {Campos}, A. and {Canning}, R.~E.~A. and {Carlstrom}, J.~E. and {Carnero Rosell}, A. and {Carrasco Kind}, M. and {Carretero}, J. and {Castander}, F.~J. and {Cawthon}, R. and {Chang}, C.~L. and {Chang}, C. and {Chaubal}, P. and {Chen}, R. and {Chiang}, H.~C. and {Choi}, A. and {Chou}, T. -L. and {Citron}, R. and {Corbett Moran}, C. and {Cordero}, J. and {Costanzi}, M. and {Crawford}, T.~M. and {Crites}, A.~T. and {da Costa}, L.~N. and {Pereira}, M.~E.~S. and {Davis}, C. and {Davis}, T.~M. and {DeRose}, J. and {Desai}, S. and {de Haan}, T. and {Diehl}, H.~T. and {Dobbs}, M.~A. and {Dodelson}, S. and {Doux}, C. and {Drlica-Wagner}, A. and {Eckert}, K. and {Elvin-Poole}, J. and {Everett}, S. and {Everett}, W. and {Ferrero}, I. and {Fert{\'e}}, A. and {Flores}, A.~M. and {Frieman}, J. and {Gallicchio}, J. and {Garc{\'\i}a-Bellido}, J. and {Gatti}, M. and {George}, E.~M. and {Giannini}, G. and {Gladders}, M.~D. and {Gruen}, D. and {Gruendl}, R.~A. and {Gupta}, N. and {Gutierrez}, G. and {Halverson}, N.~W. and {Harrison}, I. and {Hartley}, W.~G. and {Herner}, K. and {Hinton}, S.~R. and {Holder}, G.~P. and {Hollowood}, D.~L. and {Holzapfel}, W.~L. and {Honscheid}, K. and {Hrubes}, J.~D. and {Huang}, N. and {Hubmayr}, J. and {Huff}, E.~M. and {Huterer}, D. and {Irwin}, K.~D. and {James}, D.~J. and {Jarvis}, M. and {Khullar}, G. and {Kim}, K. and {Knox}, L. and {Kraft}, R. and {Krause}, E. and {Kuehn}, K. and {Kuropatkin}, N. and {K{\'e}ruzor{\'e}}, F. and {Lahav}, O. and {Lee}, A.~T. and {Leget}, P. -F. and {Li}, D. and {Lin}, H. and {Lowitz}, A. and {MacCrann}, N. and {Mahler}, G. and {Mantz}, A. and {Marshall}, J.~L. and {McCullough}, J. and {McDonald}, M. and {McMahon}, J.~J. and {Mena-Fern{\'a}ndez}, J. and {Menanteau}, F. and {Meyer}, S.~S. and {Miquel}, R. and {Montgomery}, J. and {Myles}, J. and {Natoli}, T. and {Navarro-Alsina}, A. and {Nibarger}, J.~P. and {Noble}, G.~I. and {Novosad}, V. and {Ogando}, R.~L.~C. and {Omori}, Y. and {Padin}, S. and {Pandey}, S. and {Paschos}, P. and {Patil}, S. and {Pieres}, A. and {Plazas Malag{\'o}n}, A.~A. and {Porredon}, A. and {Prat}, J. and {Pryke}, C. and {Raveri}, M. and {Reichardt}, C.~L. and {Roberson}, J. and {Rollins}, R.~P. and {Romero}, C. and {Roodman}, A. and {Ruhl}, J.~E. and {Rykoff}, E.~S. and {Saliwanchik}, B.~R. and {Salvati}, L. and {S{\'a}nchez}, C. and {Sanchez}, E. and {Sanchez Cid}, D. and {Saro}, A. and {Schaffer}, K.~K. and {Secco}, L.~F. and {Sevilla-Noarbe}, I. and {Sharon}, K. and {Sheldon}, E. and {Shin}, T. and {Sievers}, C. and {Smecher}, G. and {Smith}, M. and {Somboonpanyakul}, T. and {Sommer}, M. and {Stalder}, B. and {Stark}, A.~A. and {Stephen}, J. and {Strazzullo}, V. and {Suchyta}, E. and {Tarle}, G. and {To}, C. and {Troxel}, M.~A. and {Tucker}, C. and {Tutusaus}, I. and {Varga}, T.~N. and {Veach}, T. and {Vieira}, J.~D. and {Vikhlinin}, A. and {von der Linden}, A. and {Wang}, G. and {Weaverdyck}, N. and {Weller}, J. and {Whitehorn}, N. and {Wu}, W.~L.~K. and {Yanny}, B. and {Yefremenko}, V. and {Yin}, B. and {Young}, M. and {Zebrowski}, J.~A. and {Zhang}, Y. and {Zohren}, H. and {Zuntz}, J. and {(SPT} and {DES Collaborations)}},
        title = "{SPT clusters with DES and HST weak lensing. II. Cosmological constraints from the abundance of massive halos}",
      journal = {\prd},
         year = 2024,
        month = oct,
       volume = {110},
       number = {8},
          eid = {083510},
        pages = {083510},
          doi = {10.1103/PhysRevD.110.083510},
archivePrefix = {arXiv},
       eprint = {2401.02075},
 primaryClass = {astro-ph.CO},
       adsurl = {https://ui.adsabs.harvard.edu/abs/2024PhRvD.110h3510B}
}

@ARTICLE{PSZ2,
       author = {{Planck Collaboration} and {Ade}, P.~A.~R. and {Aghanim}, N. and {Arnaud}, M. and {Ashdown}, M. and {Aumont}, J. and {Baccigalupi}, C. and {Banday}, A.~J. and {Barreiro}, R.~B. and {Barrena}, R. and {Bartlett}, J.~G. and {Bartolo}, N. and {Battaner}, E. and {Battye}, R. and {Benabed}, K. and {Beno{\^\i}t}, A. and {Benoit-L{\'e}vy}, A. and {Bernard}, J. -P. and {Bersanelli}, M. and {Bielewicz}, P. and {Bikmaev}, I. and {B{\"o}hringer}, H. and {Bonaldi}, A. and {Bonavera}, L. and {Bond}, J.~R. and {Borrill}, J. and {Bouchet}, F.~R. and {Bucher}, M. and {Burenin}, R. and {Burigana}, C. and {Butler}, R.~C. and {Calabrese}, E. and {Cardoso}, J. -F. and {Carvalho}, P. and {Catalano}, A. and {Challinor}, A. and {Chamballu}, A. and {Chary}, R. -R. and {Chiang}, H.~C. and {Chon}, G. and {Christensen}, P.~R. and {Clements}, D.~L. and {Colombi}, S. and {Colombo}, L.~P.~L. and {Combet}, C. and {Comis}, B. and {Couchot}, F. and {Coulais}, A. and {Crill}, B.~P. and {Curto}, A. and {Cuttaia}, F. and {Dahle}, H. and {Danese}, L. and {Davies}, R.~D. and {Davis}, R.~J. and {de Bernardis}, P. and {de Rosa}, A. and {de Zotti}, G. and {Delabrouille}, J. and {D{\'e}sert}, F. -X. and {Dickinson}, C. and {Diego}, J.~M. and {Dolag}, K. and {Dole}, H. and {Donzelli}, S. and {Dor{\'e}}, O. and {Douspis}, M. and {Ducout}, A. and {Dupac}, X. and {Efstathiou}, G. and {Eisenhardt}, P.~R.~M. and {Elsner}, F. and {En{\ss}lin}, T.~A. and {Eriksen}, H.~K. and {Falgarone}, E. and {Fergusson}, J. and {Feroz}, F. and {Ferragamo}, A. and {Finelli}, F. and {Forni}, O. and {Frailis}, M. and {Fraisse}, A.~A. and {Franceschi}, E. and {Frejsel}, A. and {Galeotta}, S. and {Galli}, S. and {Ganga}, K. and {G{\'e}nova-Santos}, R.~T. and {Giard}, M. and {Giraud-H{\'e}raud}, Y. and {Gjerl{\o}w}, E. and {Gonz{\'a}lez-Nuevo}, J. and {G{\'o}rski}, K.~M. and {Grainge}, K.~J.~B. and {Gratton}, S. and {Gregorio}, A. and {Gruppuso}, A. and {Gudmundsson}, J.~E. and {Hansen}, F.~K. and {Hanson}, D. and {Harrison}, D.~L. and {Hempel}, A. and {Henrot-Versill{\'e}}, S. and {Hern{\'a}ndez-Monteagudo}, C. and {Herranz}, D. and {Hildebrandt}, S.~R. and {Hivon}, E. and {Hobson}, M. and {Holmes}, W.~A. and {Hornstrup}, A. and {Hovest}, W. and {Huffenberger}, K.~M. and {Hurier}, G. and {Jaffe}, A.~H. and {Jaffe}, T.~R. and {Jin}, T. and {Jones}, W.~C. and {Juvela}, M. and {Keih{\"a}nen}, E. and {Keskitalo}, R. and {Khamitov}, I. and {Kisner}, T.~S. and {Kneissl}, R. and {Knoche}, J. and {Kunz}, M. and {Kurki-Suonio}, H. and {Lagache}, G. and {Lamarre}, J. -M. and {Lasenby}, A. and {Lattanzi}, M. and {Lawrence}, C.~R. and {Leonardi}, R. and {Lesgourgues}, J. and {Levrier}, F. and {Liguori}, M. and {Lilje}, P.~B. and {Linden-V{\o}rnle}, M. and {L{\'o}pez-Caniego}, M. and {Lubin}, P.~M. and {Mac{\'\i}as-P{\'e}rez}, J.~F. and {Maggio}, G. and {Maino}, D. and {Mak}, D.~S.~Y. and {Mandolesi}, N. and {Mangilli}, A. and {Martin}, P.~G. and {Mart{\'\i}nez-Gonz{\'a}lez}, E. and {Masi}, S. and {Matarrese}, S. and {Mazzotta}, P. and {McGehee}, P. and {Mei}, S. and {Melchiorri}, A. and {Melin}, J. -B. and {Mendes}, L. and {Mennella}, A. and {Migliaccio}, M. and {Mitra}, S. and {Miville-Desch{\^e}nes}, M. -A. and {Moneti}, A. and {Montier}, L. and {Morgante}, G. and {Mortlock}, D. and {Moss}, A. and {Munshi}, D. and {Murphy}, J.~A. and {Naselsky}, P. and {Nastasi}, A. and {Nati}, F. and {Natoli}, P. and {Netterfield}, C.~B. and {N{\o}rgaard-Nielsen}, H.~U. and {Noviello}, F. and {Novikov}, D. and {Novikov}, I. and {Olamaie}, M. and {Oxborrow}, C.~A. and {Paci}, F. and {Pagano}, L. and {Pajot}, F. and {Paoletti}, D. and {Pasian}, F. and {Patanchon}, G. and {Pearson}, T.~J. and {Perdereau}, O. and {Perotto}, L. and {Perrott}, Y.~C. and {Perrotta}, F. and {Pettorino}, V. and {Piacentini}, F. and {Piat}, M. and {Pierpaoli}, E. and {Pietrobon}, D. and {Plaszczynski}, S. and {Pointecouteau}, E. and {Polenta}, G. and {Pratt}, G.~W. and {Pr{\'e}zeau}, G. and {Prunet}, S. and {Puget}, J. -L. and {Rachen}, J.~P. and {Reach}, W.~T. and {Rebolo}, R. and {Reinecke}, M. and {Remazeilles}, M. and {Renault}, C. and {Renzi}, A. and {Ristorcelli}, I. and {Rocha}, G. and {Rosset}, C. and {Rossetti}, M. and {Roudier}, G. and {Rozo}, E. and {Rubi{\~n}o-Mart{\'\i}n}, J.~A. and {Rumsey}, C. and {Rusholme}, B. and {Rykoff}, E.~S. and {Sandri}, M. and {Santos}, D. and {Saunders}, R.~D.~E. and {Savelainen}, M. and {Savini}, G. and {Schammel}, M.~P. and {Scott}, D. and {Seiffert}, M.~D. and {Shellard}, E.~P.~S. and {Shimwell}, T.~W. and {Spencer}, L.~D. and {Stanford}, S.~A. and {Stern}, D. and {Stolyarov}, V. and {Stompor}, R. and {Streblyanska}, A. and {Sudiwala}, R. and {Sunyaev}, R. and {Sutton}, D. and {Suur-Uski}, A. -S. and {Sygnet}, J. -F. and {Tauber}, J.~A. and {Terenzi}, L. and {Toffolatti}, L. and {Tomasi}, M. and {Tramonte}, D. and {Tristram}, M. and {Tucci}, M. and {Tuovinen}, J. and {Umana}, G. and {Valenziano}, L. and {Valiviita}, J. and {Van Tent}, B. and {Vielva}, P. and {Villa}, F. and {Wade}, L.~A. and {Wandelt}, B.~D. and {Wehus}, I.~K. and {White}, S.~D.~M. and {Wright}, E.~L. and {Yvon}, D. and {Zacchei}, A. and {Zonca}, A.},
        title = "{Planck 2015 results. XXVII. The second Planck catalogue of Sunyaev-Zeldovich sources}",
      journal = {A \& A},
         year = 2016,
        month = sep,
       volume = {594},
          eid = {A27},
        pages = {A27},
          doi = {10.1051/0004-6361/201525823},
archivePrefix = {arXiv},
       eprint = {1502.01598},
 primaryClass = {astro-ph.CO},
       adsurl = {https://ui.adsabs.harvard.edu/abs/2016A\&A...594A..27P}
}

@article{Saxena_2025,
   title={CHEX-MATE: The impact of triaxiality and orientation on Planck SZ cluster selection and weak lensing mass measurements},
   volume={700},
   ISSN={1432-0746},
   url={http://dx.doi.org/10.1051/0004-6361/202555719},
   DOI={10.1051/0004-6361/202555719},
   journal={Astronomy \&amp; Astrophysics},
   publisher={EDP Sciences},
   author={Saxena, H. and Sayers, J. and Gavidia, A. and Melin, J. -B. and Lau, E. T. and Kim, J. and Chappuis, L. and Eckert, D. and Ettori, S. and Gaspari, M. and Gastaldello, F. and Giocoli, C. and Kay, S. and Lovisari, L. and Maughan, B. and Oppizzi, F. and De Petris, M. and Pratt, G. W. and Pointecouteau, E. and Rasia, E. and Rossetti, M. and Sereno, M.},
   year={2025},
   month=aug, pages={A128} }

@article{Planck1502.01598,
	adsurl = {http://adsabs.harvard.edu/abs/2016A%26A...594A..27P},
	archiveprefix = {arXiv},
	author = {{Planck Collaboration} and {Ade}, P.~A.~R. and {Aghanim}, N. and {Arnaud}, M. and {Ashdown}, M. and {Aumont}, J. and {Baccigalupi}, C. and {Banday}, A.~J. and {Barreiro}, R.~B. and {Barrena}, R. and et al.},
	doi = {10.1051/0004-6361/201525823},
	eid = {A27},
	eprint = {1502.01598},
	journal = {\aap},
	month = sep,
	pages = {A27},
	title = {{Planck 2015 results. XXVII. The second Planck catalogue of Sunyaev-Zeldovich sources}},
	volume = 594,
	year = 2016}

@article{Hilton2009.11043,
	adsurl = {https://ui.adsabs.harvard.edu/abs/2021ApJS..253....3H},
	archiveprefix = {arXiv},
	author = {{Hilton}, M. and {Sif{\'o}n}, C. and {Naess}, S. and {Madhavacheril}, M. and {Oguri}, M. and {Rozo}, E. and {Rykoff}, E. and {Abbott}, T.~M.~C. and {Adhikari}, S. and {Aguena}, M. and {Aiola}, S. and {Allam}, S. and {Amodeo}, S. and {Amon}, A. and {Annis}, J. and {Ansarinejad}, B. and {Aros-Bunster}, C. and {Austermann}, J.~E. and {Avila}, S. and {Bacon}, D. and {Battaglia}, N. and {Beall}, J.~A. and {Becker}, D.~T. and {Bernstein}, G.~M. and {Bertin}, E. and {Bhandarkar}, T. and {Bhargava}, S. and {Bond}, J.~R. and {Brooks}, D. and {Burke}, D.~L. and {Calabrese}, E. and {Carrasco Kind}, M. and {Carretero}, J. and {Choi}, S.~K. and {Choi}, A. and {Conselice}, C. and {da Costa}, L.~N. and {Costanzi}, M. and {Crichton}, D. and {Crowley}, K.~T. and {D{\"u}nner}, R. and {Denison}, E.~V. and {Devlin}, M.~J. and {Dicker}, S.~R. and {Diehl}, H.~T. and {Dietrich}, J.~P. and {Doel}, P. and {Duff}, S.~M. and {Duivenvoorden}, A.~J. and {Dunkley}, J. and {Everett}, S. and {Ferraro}, S. and {Ferrero}, I. and {Fert{\'e}}, A. and {Flaugher}, B. and {Frieman}, J. and {Gallardo}, P.~A. and {Garc{\'\i}a-Bellido}, J. and {Gaztanaga}, E. and {Gerdes}, D.~W. and {Giles}, P. and {Golec}, J.~E. and {Gralla}, M.~B. and {Grandis}, S. and {Gruen}, D. and {Gruendl}, R.~A. and {Gschwend}, J. and {Gutierrez}, G. and {Han}, D. and {Hartley}, W.~G. and {Hasselfield}, M. and {Hill}, J.~C. and {Hilton}, G.~C. and {Hincks}, A.~D. and {Hinton}, S.~R. and {Ho}, S. -P.~P. and {Honscheid}, K. and {Hoyle}, B. and {Hubmayr}, J. and {Huffenberger}, K.~M. and {Hughes}, J.~P. and {Jaelani}, A.~T. and {Jain}, B. and {James}, D.~J. and {Jeltema}, T. and {Kent}, S. and {Knowles}, K. and {Koopman}, B.~J. and {Kuehn}, K. and {Lahav}, O. and {Lima}, M. and {Lin}, Y. -T. and {Lokken}, M. and {Loubser}, S.~I. and {MacCrann}, N. and {Maia}, M.~A.~G. and {Marriage}, T.~A. and {Martin}, J. and {McMahon}, J. and {Melchior}, P. and {Menanteau}, F. and {Miquel}, R. and {Miyatake}, H. and {Moodley}, K. and {Morgan}, R. and {Mroczkowski}, T. and {Nati}, F. and {Newburgh}, L.~B. and {Niemack}, M.~D. and {Nishizawa}, A.~J. and {Ogando}, R.~L.~C. and {Orlowski-Scherer}, J. and {Page}, L.~A. and {Palmese}, A. and {Partridge}, B. and {Paz-Chinch{\'o}n}, F. and {Phakathi}, P. and {Plazas}, A.~A. and {Robertson}, N.~C. and {Romer}, A.~K. and {Carnero Rosell}, A. and {Salatino}, M. and {Sanchez}, E. and {Schaan}, E. and {Schillaci}, A. and {Sehgal}, N. and {Serrano}, S. and {Shin}, T. and {Simon}, S.~M. and {Smith}, M. and {Soares-Santos}, M. and {Spergel}, D.~N. and {Staggs}, S.~T. and {Storer}, E.~R. and {Suchyta}, E. and {Swanson}, M.~E.~C. and {Tarle}, G. and {Thomas}, D. and {To}, C. and {Trac}, H. and {Ullom}, J.~N. and {Vale}, L.~R. and {Van Lanen}, J. and {Vavagiakis}, E.~M. and {De Vicente}, J. and {Wilkinson}, R.~D. and {Wollack}, E.~J. and {Xu}, Z. and {Zhang}, Y.},
	doi = {10.3847/1538-4365/abd023},
	eid = {3},
	eprint = {2009.11043},
	journal = {\apjs},
	month = mar,
	number = {1},
	pages = {3},
	primaryclass = {astro-ph.CO},
	title = {{The Atacama Cosmology Telescope: A Catalog of >4000 Sunyaev{\textendash}Zel{\textquoteright}dovich Galaxy Clusters}},
	volume = {253},
	year = 2021}

@article{Bulbul2402.08452,
	adsurl = {https://ui.adsabs.harvard.edu/abs/2024A\&A...685A.106B},
	archiveprefix = {arXiv},
	author = {{Bulbul}, E. and {Liu}, A. and {Kluge}, M. and {Zhang}, X. and {Sanders}, J.~S. and {Bahar}, Y.~E. and {Ghirardini}, V. and {Artis}, E. and {Seppi}, R. and {Garrel}, C. and {Ramos-Ceja}, M.~E. and {Comparat}, J. and {Balzer}, F. and {B{\"o}ckmann}, K. and {Br{\"u}ggen}, M. and {Clerc}, N. and {Dennerl}, K. and {Dolag}, K. and {Freyberg}, M. and {Grandis}, S. and {Gruen}, D. and {Kleinebreil}, F. and {Krippendorf}, S. and {Lamer}, G. and {Merloni}, A. and {Migkas}, K. and {Nandra}, K. and {Pacaud}, F. and {Predehl}, P. and {Reiprich}, T.~H. and {Schrabback}, T. and {Veronica}, A. and {Weller}, J. and {Zelmer}, S.},
	doi = {10.1051/0004-6361/202348264},
	eid = {A106},
	eprint = {2402.08452},
	journal = {\aap},
	month = may,
	pages = {A106},
	primaryclass = {astro-ph.CO},
	title = {{The SRG/eROSITA All-Sky Survey. The first catalog of galaxy clusters and groups in the Western Galactic Hemisphere}},
	volume = {685},
	year = 2024}

@article{DES2002.11124,
	adsurl = {https://ui.adsabs.harvard.edu/abs/2020PhRvD.102b3509A},
	archiveprefix = {arXiv},
	author = {{Abbott}, T.~M.~C. and {Aguena}, M. and {Alarcon}, A. and {Allam}, S. and {Allen}, S. and {Annis}, J. and {Avila}, S. and {Bacon}, D. and {Bechtol}, K. and {Bermeo}, A. and {Bernstein}, G.~M. and {Bertin}, E. and {Bhargava}, S. and {Bocquet}, S. and {Brooks}, D. and {Brout}, D. and {Buckley-Geer}, E. and {Burke}, D.~L. and {Carnero Rosell}, A. and {Carrasco Kind}, M. and {Carretero}, J. and {Castander}, F.~J. and {Cawthon}, R. and {Chang}, C. and {Chen}, X. and {Choi}, A. and {Costanzi}, M. and {Crocce}, M. and {da Costa}, L.~N. and {Davis}, T.~M. and {De Vicente}, J. and {DeRose}, J. and {Desai}, S. and {Diehl}, H.~T. and {Dietrich}, J.~P. and {Dodelson}, S. and {Doel}, P. and {Drlica-Wagner}, A. and {Eckert}, K. and {Eifler}, T.~F. and {Elvin-Poole}, J. and {Estrada}, J. and {Everett}, S. and {Evrard}, A.~E. and {Farahi}, A. and {Ferrero}, I. and {Flaugher}, B. and {Fosalba}, P. and {Frieman}, J. and {Garc{\'\i}a-Bellido}, J. and {Gatti}, M. and {Gaztanaga}, E. and {Gerdes}, D.~W. and {Giannantonio}, T. and {Giles}, P. and {Grandis}, S. and {Gruen}, D. and {Gruendl}, R.~A. and {Gschwend}, J. and {Gutierrez}, G. and {Hartley}, W.~G. and {Hinton}, S.~R. and {Hollowood}, D.~L. and {Honscheid}, K. and {Hoyle}, B. and {Huterer}, D. and {James}, D.~J. and {Jarvis}, M. and {Jeltema}, T. and {Johnson}, M.~W.~G. and {Johnson}, M.~D. and {Kent}, S. and {Krause}, E. and {Kron}, R. and {Kuehn}, K. and {Kuropatkin}, N. and {Lahav}, O. and {Li}, T.~S. and {Lidman}, C. and {Lima}, M. and {Lin}, H. and {MacCrann}, N. and {Maia}, M.~A.~G. and {Mantz}, A. and {Marshall}, J.~L. and {Martini}, P. and {Mayers}, J. and {Melchior}, P. and {Mena-Fern{\'a}ndez}, J. and {Menanteau}, F. and {Miquel}, R. and {Mohr}, J.~J. and {Nichol}, R.~C. and {Nord}, B. and {Ogando}, R.~L.~C. and {Palmese}, A. and {Paz-Chinch{\'o}n}, F. and {Plazas}, A.~A. and {Prat}, J. and {Rau}, M.~M. and {Romer}, A.~K. and {Roodman}, A. and {Rooney}, P. and {Rozo}, E. and {Rykoff}, E.~S. and {Sako}, M. and {Samuroff}, S. and {S{\'a}nchez}, C. and {Sanchez}, E. and {Saro}, A. and {Scarpine}, V. and {Schubnell}, M. and {Scolnic}, D. and {Serrano}, S. and {Sevilla-Noarbe}, I. and {Sheldon}, E. and {Smith}, J. Allyn. and {Smith}, M. and {Suchyta}, E. and {Swanson}, M.~E.~C. and {Tarle}, G. and {Thomas}, D. and {To}, C. and {Troxel}, M.~A. and {Tucker}, D.~L. and {Varga}, T.~N. and {von der Linden}, A. and {Walker}, A.~R. and {Wechsler}, R.~H. and {Weller}, J. and {Wilkinson}, R.~D. and {Wu}, H. and {Yanny}, B. and {Zhang}, Y. and {Zhang}, Z. and {Zuntz}, J. and {DES Collaboration}},
	doi = {10.1103/PhysRevD.102.023509},
	eid = {023509},
	eprint = {2002.11124},
	journal = {\prd},
	month = jul,
	number = {2},
	pages = {023509},
	primaryclass = {astro-ph.CO},
	title = {{Dark Energy Survey Year 1 Results: Cosmological constraints from cluster abundances and weak lensing}},
	volume = {102},
	year = 2020}

@article{Mantz1901.10522,
	adsurl = {https://ui.adsabs.harvard.edu/abs/2019MNRAS.485.4863M},
	archiveprefix = {arXiv},
	author = {{Mantz}, Adam B.},
	doi = {10.1093/mnras/stz320},
	eprint = {1901.10522},
	journal = {\mnras},
	month = {Jun},
	number = {4},
	pages = {4863-4872},
	primaryclass = {astro-ph.IM},
	title = {{Coping with selection effects: a Primer on regression with truncated data}},
	volume = {485},
	year = {2019}}

@article{Ebeling0009101,
	adsurl = {http://adsabs.harvard.edu/cgi-bin/nph-bib_query?bibcode=2001ApJ...553..668E&db_key=AST},
	author = {{Ebeling}, H. and {Edge}, A.~C. and {Henry}, J.~P.},
	doi = {10.1086/320958},
	eprint = {astro-ph/0009101},
	journal = {\apj},
	month = jun,
	pages = {668-676},
	title = {{MACS: A Quest for the Most Massive Galaxy Clusters in the Universe}},
	volume = 553,
	year = 2001}

@article{Bleem1409.0850,
	adsurl = {http://adsabs.harvard.edu/abs/2015ApJS..216...27B},
	archiveprefix = {arXiv},
	author = {{Bleem}, L.~E. and {Stalder}, B. and {de Haan}, T. and {Aird}, K.~A. and {Allen}, S.~W. and {Applegate}, D.~E. and {Ashby}, M.~L.~N. and {Bautz}, M. and {Bayliss}, M. and {Benson}, B.~A. and {Bocquet}, S. and {Brodwin}, M. and {Carlstrom}, J.~E. and {Chang}, C.~L. and {Chiu}, I. and {Cho}, H.~M. and {Clocchiatti}, A. and {Crawford}, T.~M. and {Crites}, A.~T. and {Desai}, S. and {Dietrich}, J.~P. and {Dobbs}, M.~A. and {Foley}, R.~J. and {Forman}, W.~R. and {George}, E.~M. and {Gladders}, M.~D. and {Gonzalez}, A.~H. and {Halverson}, N.~W. and {Hennig}, C. and {Hoekstra}, H. and {Holder}, G.~P. and {Holzapfel}, W.~L. and {Hrubes}, J.~D. and {Jones}, C. and {Keisler}, R. and {Knox}, L. and {Lee}, A.~T. and {Leitch}, E.~M. and {Liu}, J. and {Lueker}, M. and {Luong-Van}, D. and {Mantz}, A. and {Marrone}, D.~P. and {McDonald}, M. and {McMahon}, J.~J. and {Meyer}, S.~S. and {Mocanu}, L. and {Mohr}, J.~J. and {Murray}, S.~S. and {Padin}, S. and {Pryke}, C. and {Reichardt}, C.~L. and {Rest}, A. and {Ruel}, J. and {Ruhl}, J.~E. and {Saliwanchik}, B.~R. and {Saro}, A. and {Sayre}, J.~T. and {Schaffer}, K.~K. and {Schrabback}, T. and {Shirokoff}, E. and {Song}, J. and {Spieler}, H.~G. and {Stanford}, S.~A. and {Staniszewski}, Z. and {Stark}, A.~A. and {Story}, K.~T. and {Stubbs}, C.~W. and {Vanderlinde}, K. and {Vieira}, J.~D. and {Vikhlinin}, A. and {Williamson}, R. and {Zahn}, O. and {Zenteno}, A.},
	doi = {10.1088/0067-0049/216/2/27},
	eid = {27},
	eprint = {1409.0850},
	journal = {\apjs},
	month = feb,
	pages = {27},
	title = {{Galaxy Clusters Discovered via the Sunyaev-Zel'dovich Effect in the 2500-Square-Degree SPT-SZ Survey}},
	volume = 216,
	year = 2015}

@article{Sayers1209.5129,
	adsurl = {http://adsabs.harvard.edu/abs/2013ApJ...764..152S},
	archiveprefix = {arXiv},
	author = {{Sayers}, J. and {Mroczkowski}, T. and {Czakon}, N.~G. and {Golwala}, S.~R. and {Mantz}, A. and {Ameglio}, S. and {Downes}, T.~P. and {Koch}, P.~M. and {Lin}, K.-Y. and {Molnar}, S.~M. and {Moustakas}, L. and {Muchovej}, S.~J.~C. and {Pierpaoli}, E. and {Shitanishi}, J.~A. and {Siegel}, S. and {Umetsu}, K.},
	doi = {10.1088/0004-637X/764/2/152},
	eid = {152},
	eprint = {1209.5129},
	journal = {\apj},
	month = feb,
	pages = {152},
	primaryclass = {astro-ph.CO},
	title = {{The Contribution of Radio Galaxy Contamination to Measurements of the Sunyaev-Zel'dovich Decrement in Massive Galaxy Clusters at 140 GHz with Bolocam}},
	volume = 764,
	year = 2013}

@ARTICLE{Gupta2017,
       author = {{Gupta}, N. and {Saro}, A. and {Mohr}, J.~J. and {Benson}, B.~A. and {Bocquet}, S. and {Capasso}, R. and {Carlstrom}, J.~E. and {Chiu}, I. and {Crawford}, T.~M. and {de Haan}, T. and {Dietrich}, J.~P. and {Gangkofner}, C. and {Holzapfel}, W.~L. and {McDonald}, M. and {Rapetti}, D. and {Reichardt}, C.~L.},
        title = "{High-frequency cluster radio galaxies: luminosity functions and implications for SZE-selected cluster samples}",
      journal = {\mnras},
         year = 2017,
        month = may,
       volume = {467},
       number = {3},
        pages = {3737-3750},
          doi = {10.1093/mnras/stx095},
archivePrefix = {arXiv},
       eprint = {1605.05329},
 primaryClass = {astro-ph.CO},
       adsurl = {https://ui.adsabs.harvard.edu/abs/2017MNRAS.467.3737G}
}

@ARTICLE{Mo_AGN_mass_clusters,
       author = {{Mo}, Wenli and {Gonzalez}, Anthony and {Brodwin}, Mark and {Decker}, Bandon and {Eisenhardt}, Peter and {Moravec}, Emily and {Stanford}, S.~A. and {Stern}, Daniel and {Wylezalek}, Dominika},
        title = "{The Massive and Distant Clusters of WISE Survey. VIII. Radio Activity in Massive Galaxy Clusters}",
      journal = {\apj},
         year = 2020,
        month = oct,
       volume = {901},
       number = {2},
          eid = {131},
        pages = {131},
          doi = {10.3847/1538-4357/abb08d},
archivePrefix = {arXiv},
       eprint = {2008.07686},
 primaryClass = {astro-ph.GA},
       adsurl = {https://ui.adsabs.harvard.edu/abs/2020ApJ...901..131M}
}

@article{Eckert_2019,
	author = {{Eckert, D.} and {Ghirardini, V.} and {Ettori, S.} and {Rasia, E.} and {Biffi, V.} and {Pointecouteau, E.} and {Rossetti, M.} and {Molendi, S.} and {Vazza, F.} and {Gastaldello, F.} and {Gaspari, M.} and {De Grandi, S.} and {Ghizzardi, S.} and {Bourdin, H.} and {Tchernin, C.} and {Roncarelli, M.}},
	title = {Non-thermal pressure support in X-COP galaxy clusters},
	DOI= "10.1051/0004-6361/201833324",
	url= "https://doi.org/10.1051/0004-6361/201833324",
	journal = {A\&A},
	year = 2019,
	volume = 621,
	pages = "A40",
}

@article{Hlavacek_Larrondo_2012,
   title={On the hunt for ultramassive black holes in brightest cluster galaxies: UMBHs in BCGs},
   volume={424},
   ISSN={0035-8711},
   url={http://dx.doi.org/10.1111/j.1365-2966.2012.21187.x},
   DOI={10.1111/j.1365-2966.2012.21187.x},
   number={1},
   journal={Monthly Notices of the Royal Astronomical Society},
   publisher={Oxford University Press (OUP)},
   author={Hlavacek-Larrondo, J. and Fabian, A. C. and Edge, A. C. and Hogan, M. T.},
   year={2012},
   month=jun, pages={224–231} }

@ARTICLE{Hogan15b,
       author = {{Hogan}, M.~T. and {Edge}, A.~C. and {Geach}, J.~E. and {Grainge}, K.~J.~B. and {Hlavacek-Larrondo}, J. and {Hovatta}, T. and {Karim}, A. and {McNamara}, B.~R. and {Rumsey}, C. and {Russell}, H.~R. and {Salom{\'e}}, P. and {Aller}, H.~D. and {Aller}, M.~F. and {Benford}, D.~J. and {Fabian}, A.~C. and {Readhead}, A.~C.~S. and {Sadler}, E.~M. and {Saunders}, R.~D.~E.},
        title = "{High radio-frequency properties and variability of brightest cluster galaxies}",
      journal = {\mnras},
         year = 2015,
        month = oct,
       volume = {453},
       number = {2},
        pages = {1223-1240},
          doi = {10.1093/mnras/stv1518},
archivePrefix = {arXiv},
       eprint = {1507.03022},
 primaryClass = {astro-ph.GA},
       adsurl = {https://ui.adsabs.harvard.edu/abs/2015MNRAS.453.1223H}
}

@ARTICLE{Timmerman_2023,
       author = {{Timmerman}, R. and {van Weeren}, R.~J. and {Callingham}, J.~R. and {Cotton}, W.~D. and {Perley}, R. and {Morabito}, L.~K. and {Gizani}, N.~A.~B. and {Bridle}, A.~H. and {O'Dea}, C.~P. and {Baum}, S.~A. and {Tremblay}, G.~R. and {Kharb}, P. and {Kassim}, N.~E. and {R{\"o}ttgering}, H.~J.~A. and {Botteon}, A. and {Sweijen}, F. and {Tasse}, C. and {Br{\"u}ggen}, M. and {Moldon}, J. and {Shimwell}, T. and {Brunetti}, G.},
        title = "{Origin of the ring structures in Hercules A. Sub-arcsecond 144 MHz to 7 GHz observations}",
      journal = {\aap},
         year = 2022,
        month = feb,
       volume = {658},
          eid = {A5},
        pages = {A5},
          doi = {10.1051/0004-6361/202140880},
archivePrefix = {arXiv},
       eprint = {2108.07287},
 primaryClass = {astro-ph.GA},
       adsurl = {https://ui.adsabs.harvard.edu/abs/2022A\&A...658A...5T}
}

@article{Choi_2020,
   title={The Atacama Cosmology Telescope: a measurement of the Cosmic Microwave Background power spectra at 98 and 150 GHz},
   volume={2020},
   ISSN={1475-7516},
   url={http://dx.doi.org/10.1088/1475-7516/2020/12/045},
   DOI={10.1088/1475-7516/2020/12/045},
   number={12},
   journal={Journal of Cosmology and Astroparticle Physics},
   publisher={IOP Publishing},
   author={Choi, Steve K. and Hasselfield, Matthew and Ho, Shuay-Pwu Patty and Koopman, Brian and Lungu, Marius and Abitbol, Maximilian H. and Addison, Graeme E. and Ade, Peter A. R. and Aiola, Simone and Alonso, David and Amiri, Mandana and Amodeo, Stefania and Angile, Elio and Austermann, Jason E. and Baildon, Taylor and Battaglia, Nick and Beall, James A. and Bean, Rachel and Becker, Daniel T. and Bond, J Richard and Bruno, Sarah Marie and Calabrese, Erminia and Calafut, Victoria and Campusano, Luis E. and Carrero, Felipe and Chesmore, Grace E. and Cho, Hsiao-mei and Clark, Susan E. and Cothard, Nicholas F. and Crichton, Devin and Crowley, Kevin T. and Darwish, Omar and Datta, Rahul and Denison, Edward V. and Devlin, Mark J. and Duell, Cody J. and Duff, Shannon M. and Duivenvoorden, Adriaan J. and Dunkley, Jo and Dünner, Rolando and Essinger-Hileman, Thomas and Fankhanel, Max and Ferraro, Simone and Fox, Anna E. and Fuzia, Brittany and Gallardo, Patricio A. and Gluscevic, Vera and Golec, Joseph E. and Grace, Emily and Gralla, Megan and Guan, Yilun and Hall, Kirsten and Halpern, Mark and Han, Dongwon and Hargrave, Peter and Henderson, Shawn and Hensley, Brandon and Hill, J. Colin and Hilton, Gene C. and Hilton, Matt and Hincks, Adam D. and Hložek, Renée and Hubmayr, Johannes and Huffenberger, Kevin M. and Hughes, John P. and Infante, Leopoldo and Irwin, Kent and Jackson, Rebecca and Klein, Jeff and Knowles, Kenda and Kosowsky, Arthur and Lakey, Vincent and Li, Dale and Li, Yaqiong and Li, Zack and Lokken, Martine and Louis, Thibaut and MacInnis, Amanda and Madhavacheril, Mathew and Maldonado, Felipe and Mallaby-Kay, Maya and Marsden, Danica and Maurin, Loïc and McMahon, Jeff and Menanteau, Felipe and Moodley, Kavilan and Morton, Tim and Naess, Sigurd and Namikawa, Toshiya and Nati, Federico and Newburgh, Laura and Nibarger, John P. and Nicola, Andrina and Niemack, Michael D. and Nolta, Michael R. and Orlowski-Sherer, John and Page, Lyman A. and Pappas, Christine G. and Partridge, Bruce and Phakathi, Phumlani and Prince, Heather and Puddu, Roberto and Qu, Frank J. and Rivera, Jesus and Robertson, Naomi and Rojas, Felipe and Salatino, Maria and Schaan, Emmanuel and Schillaci, Alessandro and Schmitt, Benjamin L. and Sehgal, Neelima and Sherwin, Blake D. and Sierra, Carlos and Sievers, Jon and Sifon, Cristobal and Sikhosana, Precious and Simon, Sara and Spergel, David N. and Staggs, Suzanne T. and Stevens, Jason and Storer, Emilie and Sunder, Dhaneshwar D. and Switzer, Eric R. and Thorne, Ben and Thornton, Robert and Trac, Hy and Treu, Jesse and Tucker, Carole and Vale, Leila R. and Engelen, Alexander Van and Lanen, Jeff Van and Vavagiakis, Eve M. and Wagoner, Kasey and Wang, Yuhan and Ward, Jonathan T. and Wollack, Edward J. and Xu, Zhilei and Zago, Fernando and Zhu, Ningfeng},
   year={2020},
   month=dec, pages={045–045} }

@article{Arnaud_2010,
	author = {{Arnaud, M.} and {Pratt, G. W.} and {Piffaretti, R.} and {Böhringer, H.} and {Croston, J. H.} and {Pointecouteau, E.}},
	title = {The universal galaxy cluster pressure profile from a representative sample of nearby systems (REXCESS) and the YSZ – M500 relation },
	DOI= "10.1051/0004-6361/200913416",
	url= "https://doi.org/10.1051/0004-6361/200913416",
	journal = {A\&A},
	year = 2010,
	volume = 517,
	pages = "A92",
	month = "",
}

@misc{AGNfitter,
      title={AGNfitter-rx: Modelling the radio-to-X-ray SEDs of AGNs}, 
      author={L. N. Martínez-Ramírez and G. Calistro Rivera and Elisabeta Lusso and F. E. Bauer and Emanuele Nardini and Johannes Buchner and Michael J. I. Brown and Juan C. B. Pineda and Matthew J. Temple and Manda Banerji and M. Stalevski and Joseph F. Hennawi},
      year={2024},
      eprint={2405.12111},
      archivePrefix={arXiv},
      primaryClass={astro-ph.GA},
      url={https://arxiv.org/abs/2405.12111}, 
}

@ARTICLE{SED_fitting,
       author = {{Tingay}, S.~J. and {de Kool}, M.},
        title = "{An Investigation of Synchrotron Self-absorption and Free-Free Absorption Models in Explanation of the Gigahertz-peaked Spectrum of PKS 1718-649}",
      journal = {\aj},
         year = 2003,
        month = aug,
       volume = {126},
       number = {2},
        pages = {723-733},
          doi = {10.1086/376600},
       adsurl = {https://ui.adsabs.harvard.edu/abs/2003AJ....126..723T}
}

@article{Muchovej_2007,
doi = {10.1086/511971},
url = {https://doi.org/10.1086/511971},
year = {2007},
month = {jul},
publisher = {},
volume = {663},
number = {2},
pages = {708},
author = {Muchovej, Stephen and Mroczkowski, Tony and Carlstrom, John E. and Cartwright, John and Greer, Christopher and Hennessy, Ryan and Loh, Michael and Pryke, Clem and Reddall, Ben and Runyan, Marcus and Sharp, Matthew and Hawkins, David and Lamb, James W. and Woody, David and Joy, Marshall and Leitch, Erik M. and Miller, Amber D.},
title = {Observations of High-Redshift X-Ray Selected Clusters with the Sunyaev-Zel’dovich Array},
journal = {The Astrophysical Journal}
}

@misc{Hart_RXJ1651,
      title={Evolution of Substructure in Galaxy Clusters as Observed in X-Rays}, 
      author={Brian C. Hart},
      year={2008},
      eprint={0801.4093},
      archivePrefix={arXiv},
      primaryClass={astro-ph},
      url={https://arxiv.org/abs/0801.4093}, 
}
\bibliographystyle{aasjournalv7}



\end{document}